# Robust Time-Series Retrieval Using Probabilistic Adaptive Segmental Alignment

**Shahriar Shariat and Vladimir Pavlovic**

**Abstract** Traditional pairwise sequence alignment is based on matching individual samples from two sequences, under time monotonicity constraints. However, in many application settings matching subsequences (segments) instead of individual samples may bring in additional robustness to noise or local non-causal perturbations. This paper presents an approach to segmental sequence alignment that jointly segments and aligns two sequences, generalizing the traditional per-sample alignment. To accomplish this task, we introduce a distance metric between segments based on average pairwise distances and then present a modified pair-HMM (PHMM) that incorporates the proposed distance metric to solve the joint segmentation and alignment task. We also propose a relaxation to our model that improves the computational efficiency of the generic segmental PHMM. Our results demonstrate that this new measure of sequence similarity can lead to improved classification performance, while being resilient to noise, on a variety of sequence retrieval problems, from EEG to motion sequence classification.

## 1 Introduction

Many problems in data analytics today critically depend on comparison and retrieval of time-series data, such as the stock market prices, medical signals, or moving object trajectories. The non-Euclidean nature of the space of sequences has given rise to domain-specific approaches and algorithms for general analytics tasks, including indexing, classification and clustering of time-series or sequences

Shahriar Shariat
Rutgers University, NJ 08854
E-mail: sshariat@cs.rutgers.edu
*Current affiliation:* Turn Inc., E-mail: sshariat@turn.com.

Vladimir Pavlovic
Rutgers University, NJ 08854
E-mail: Vladimir@cs.rutgers.edu

Keogh (2006)); Aghabozorgi et al (2015)). Asserting the pairwise sequence similarity is at the core of these tasks Morse and Patel (2007)); Pree et al (2014)). A family of alignment algorithms accomplishes this by measuring similarities between pairs of samples across two sequences and matching them under monotonicity (i.e., temporal ordering) constraints.

Dynamic time warping (DTW) c.f., Berndt and Clifford (1994)) is a common computational technique to tackle the problem of measuring the pairwise sequence similarity Ding et al (2008)). DTW alignment algorithms are based on pairing of individual sequence samples. That is, a sample at time $i$ in sequence $X$ is typically matched with only one other sample at time $j$ in sequence $Y$, while guaranteeing monotonic ordering, i.e., that a subsequent sample $X_{i+1}$ in one sequence could not be simultaneously matched with a preceding sample $Y_{j-1}$ in the second sequence.

Another category of similarity measurement methods are designed based on the edit distance algorithm Atallah and Fox (1998)). Examples of such approaches include Longest Common Sub-Sequence (LCSS) Andre-Jonsson and Badal (1997)); Vlachos et al (2002)), Edit distance with Real Penalty (ERP) Chen and Ng (2004)) and Edit Distance for Real Sequences (EDR) proposed in Chen and zsu (2005)). These algorithms compare the pairwise distance of two points against a threshold (pre-defined or variable) and revert the problem back to the original edit distance problem. A comprehensive review that evaluates many similarity measures across a range of benchmark tasks in Ding et al (2008)) concludes that no single algorithm consistently outperform others. Nevertheless, DTW itself was demonstrated to remain a competitive baseline, particularly in instances of noise-free or low noise time-series.

One consequence of DTW's essential reliance on comparison of pairs of individual time-series samples is its, as well as many of its derivatives', sensitivity to noise Shariat and Pavlovic (2011)); Ye and Keogh (2009)); Vlachos et al (2002)); Zakaria et al (2015)). We have observed that in the presence of significant noise, edit-distance based methods outperforms DTW. If such noise is to be removed by means of preprocessing, DTW-based comparison could again become a stable measure of sequence similarity. However, effective noise removal if often domain-specific, may require adaptation to follow the changing sequence dynamics, and, most critically, typically considers denoising of one sequence outside the context of the sequence it is being compared to. As a consequence, the denoising becomes decoupled from the process of measuring sequence similarity and, in turn, the retrieval or classification end-goals.

DTW-family algorithms are also constrained to preserve the time mononoonicity. In case of non-causal signals where local ordering of samples can change, such as the EEG time series de Munck et al (2007)) or signals with general random time delays Blaum and Bruck (1994)), DTW is not able to any more yield reliable pairwise similarity measures. Finally, in many applications, such as video segmentation, one might be interested in not only calculating the similarity but also retrieving the locally similar segments of the contrasting sequences Shariat and Pavlovic (2013)), which may constitute meaningful units of local similarity. With its focus on per-sample alignments, DTW cannot inherently produce such delineation.

As a consequence, to achieve both resilience against multiple types of noise and recover similar segments, it is reasonable to establish pairing between groups of points in contrasting sequences. That is, one may seek to match a temporal segment (contiguous subsequence) $X_{i:i+m} = [x_i, \ldots, x_{i+m}]$ to another segment of the contrasting sequence, $Y_{j:j+n} = [y_j, \ldots, y_{j+n}]$ as the basic units employed in full matching of the two sequences. In other words, the process of establishing pairwise sequence similarity needs to involve *simultaneous* segmentation of the two sequences being compared as well as their comparison that depends on the identified segments, while now satisfying the monotonicity in the order of paired segments rather than individual paired samples. We call this the adaptive segmental alignment task.

In Shariat and Pavlovic (2011)) the authors proposed an approach, based on canonical correlation analysis (CCA), to handle this segmental alignment. The objective function (IsoCCA) is constrained properly to impose time monotonicity over segments. Although the results show strong resilience to noise, the objective does not provide a proper metric between the segments. This can cause the resulting segments to be unnecessarily short. Furthermore, the non-convexity of IsoCCA objective makes it increasingly sensitive to initial segmentation and model parameter choices. Another recent work, Ryoo (2011)), proposes to find the best matching segments of the two sequences based on a probabilistic model. However, the algorithm does not handle gaps/insertions and, hence, does not consider a complete alignment model. Moreover, the author suggests empirically fixing all segment lengths, with the approach lacking clear means to handle data-driven segments. In practice, however, variable and data-adapted segments result in more robust alignments.

In Ye and Keogh (2009)), L. Ye and E. Keogh propose a method (shapelet) to discover a common subsequence between a class of time-series and take that as a class representative. This way, they overcome possible scattered noise processes that could contaminate the classification procedure. In contrast, our approach is not a motif discovery algorithm and is essentially an alignment algorithm that enhances the pairwise similarity of two sequences through discovery and matching of similar segments.

In this paper we propose a complete segmental alignment framework to address the deficiencies of prior segmental sequence comparison approaches. Specifically, the new contributions of this work are:

- We propose a distance metric based on average pair-wise distances suitable for measuring similarity between two segments, and aimed at segmental sequence alignment.
- Based on the proposed distance metric we develop a probabilistic alignment model by extending the traditional pair-HMM formalism.
- We propose a relaxation to the original model and use bounding techniques to reduce the computation time necessary to optimize the model and, hence, evaluate the pairwise segmental alignments.

Since the order of points is ignored within a segment, the algorithm is able to handle non-causal signals. Segment matching is particularly interesting in action recognition scenarios considering that actions can be easily divided in sub-actions (for example walking with long and short strides). Furthermore, the direction of the progress is not important within each segment and thus two actions that are performed in different directions might still, as desired, exhibit high similarity. The properties of the new similarity metric make it very resilient to noise and thus applicable to situations where the conventional noise removal techniques combined with traditional alignment algorithms fail to produce a reliable similarity measure. In such cases, our method combines the properties of an adaptive filter and an alignment algorithm, leading to more robust estimate of the similarity of contrasting sequences.

Through extensive experiments we show that the proposed segmental sequence alignment and similarity measure can lead to improved classification results on benchmark sequence classification tasks, classification of non-causal EEG signals, and recognition of activities from human motion data. This contrasts the often inconsistent performance of the competing approaches that either lack the ability to match segments instead of individual samples, or assume fixed, non-adaptive segmentation.

The paper is organized as follows: in Section 2 we discuss the metric property of IsoCCA and construct our segmental metric. In Section 3 the proposed model is discussed in detail. Section 4 introduces the relaxed model for reduced computational time. In Section 5 experimental results is presented followed by Section 6 that concludes the paper with the discussion of our findings and some suggestions for future work.

## 2 Segment Matching Metric

Central to any alignment algorithm is the distance metric between two points (or segments in our case). DTW, typically, assumes Euclidean distance between contrasting entities. Edit-distance-based methods, such as LCSS and EDR, measure the Euclidean or L1 distance of two points and test it against a threshold. The aforementioned algorithms are still based on the point-wise comparison of the sequences.

In Shariat and Pavlovic (2011)) the authors proposed a segmental alignment method based on CCA, *i.e.*, IsoCCA. Despite promising results, the proposed framework does not provide a proper metric between the segments. The reason for that lies in the fact that IsoCCA works by effectively finding the closet points of the convex hulls of the two segments of points. This results in a non-metric because the triangular inequality does not hold. Moreover in the case of overlapping convex hulls, their distance is zero even though the size of the common area can be very small resulting in unnecessarily small segments.

In some applications, as illustrated in Section 1, one is interested in matching unordered small segments of points where permutation of the points is not a matter of concern. In addition to insensitivity to the permutation, we seek to find a distance metric that suppresses the noise and is efficient to compute. Many distance metrics have been proposed to measure the distance between sets, c.f., Woznica et al (2006)). Often the proposed distances are based on non-linear functions (Hausdorff, for instance), which are computationally intensive. Moreover, Hausdorff-type distances can be highly insensitive to the content of the contrasting sets, focusing instead on the boundary cases. Kernels proposed on sets Kondor (2003)) are not also suitable when the set of points is small and therefore, in practice the estimated distribution is inaccurate. In the following we propose a distance based on average pair-wise distances.

Formally, for two sets of points $\mathcal{X}$ and $\mathcal{Y}$, we consider

$$d(\mathcal{X},\mathcal{Y}) = \frac{1}{|\mathcal{X}||\mathcal{Y}|}\sum_{x_i\in\mathcal{X}}\sum_{y_j\in\mathcal{Y}}\|x_i - y_j\|_n, \quad (1)$$

where $\|.\|_n$ is a convex norm between two points. It is trivial to show $d(\mathcal{X},\mathcal{Y}) \geq 0$ and $d(\mathcal{X},\mathcal{Y}) = d(\mathcal{Y},\mathcal{X})$. It is also straightforward to prove that (1) has the triangular property given the convexity of the norms. Equation (1) needs to be slightly modified to have definiteness property (i.e $d(x,y) = 0 \iff x = y$).

$$\mathcal{D}(\mathcal{X},\mathcal{Y}) = \frac{1}{|\mathcal{X}\cup\mathcal{Y}|}\left(\frac{1}{|\mathcal{X}|}\sum_{x_i\in\mathcal{X}}\sum_{y_i\in(\mathcal{Y}\setminus\mathcal{X})}\|x_i - y_j\|_n + \frac{1}{|\mathcal{Y}|}\sum_{x_i\in(\mathcal{X}\setminus\mathcal{Y})}\sum_{y_i\in\mathcal{Y}}\|x_i - y_j\|_n\right). \quad (2)$$

Equation (2) is symmetric, non-negative and definite due to empty sums in case of equality of $\mathcal{X}$ and $\mathcal{Y}$. To prove that (2) has triangular property, one can partition $(\mathcal{D}(\mathcal{X},\mathcal{Y}) + \mathcal{D}(\mathcal{Y},\mathcal{Z}) - \mathcal{D}(\mathcal{X},\mathcal{Z})) \geq 0$ into disjoint sets and observe that given triangular property of (1), the required inequality holds for (2). Note that in case of $\mathcal{X}\cap\mathcal{Y} = \emptyset$, (2) reduces to (1). In practice, any sampling is prone to measurement error and one needs to compare all pair-wise distances against that error. This emphasizes the importance of definiteness property imposed by (2) even for real-valued signals. We will show in the experimental results that even though the ordering of samples is not preserved within a short segment when modeled as a set, the proposed metric can be used for general purpose alignment. The metric also exhibits invariance to arbitrary temporal permutations. This can be beneficial for non-causal sequences that arise from random delays (e.g., EEG). However, it can also be desirable in video retrieval settings when, for instance, the direction of an activity is not a concern. In the experiments we will demonstrate that this metric is resilient to noise when incorporated into an alignment algorithm. In Section 3 we demonstrate how it can be computed efficiently.

## 3 Segmental Pair-HMM (SPHMM)

In this section we describe the details of our alignment models and algorithm. We first describe the basis of our model, a variation of Pair-HMM and its formalism. The inference algorithm works by, essentially, fixing the segment size in each step and then dynamically adjusting it to recover the best segments. We reveal the computational techniques, based on Viterbi decoding, that make this task efficient. We also propose a forward algorithm whose primary aim is to yield the similarity measure of interest without explicitly determining the segments, an approach sufficient for e.g., classification tasks. Finally, we present the SPHMM learning methodology, based on the proposed inference algorithm.

The Pair HMM, introduced by Durbin et al (1997)), can be seen as a probabilistic model defined on pairs of sequences $(X, Y)$ that aims to describe their joint likelihood, $P(X, Y|alignment)$. As shown in Figure 3, PHMM has three states:

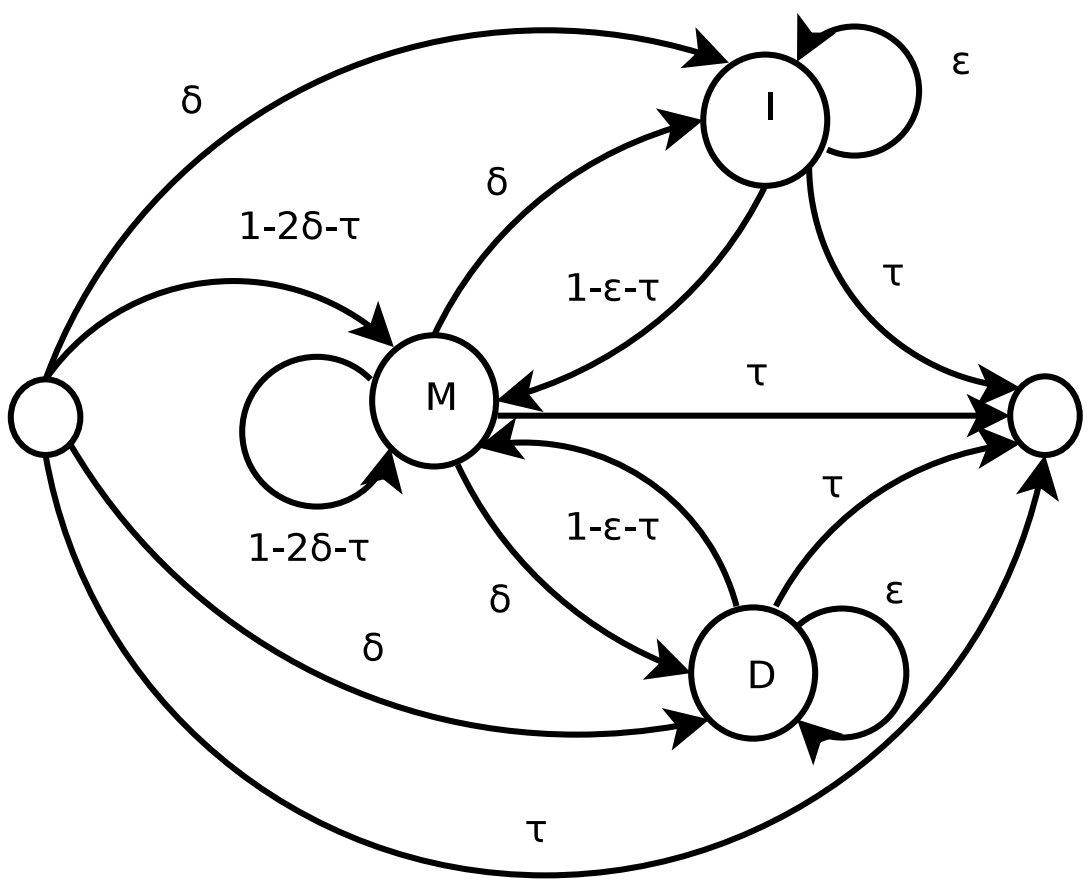

**Fig. 1** Segmental Pair-HMM state-transition diagram

$M$ for match, $I$ for insertion and $D$ for deletion. Given two sequences of observations $X$ and $Y$ with $n$ and $m$ samples, respectively, the match state emits a pair of samples $(x, y)$ $x \in X$, $y \in Y$. Insertion and deletion states emit $(x, -)$ and $(-, y)$ respectively where $-$ stands for a gap. This model implements an affine gap penalty which is more general than constant gap penalty typically used in DTW.

In the following we add the notion of segmentation to the pair-HMM formalism. To define the segmentation structure consider a sequence $X = (x_1, x_2, \dots x_n)$ of length $n$. A segment $X_{b:e}$, a contiguous subsequence of $X$, is defined such that $X_{b:e} = (x_b, x_{b+1}, \dots, x_e)$. Equivalently, the segment is defined by segment indexes $s = (b, b+1, \dots e)$. We consider non-overlapping and tight segments over $X$. That is, a complete segmentation of $X$ is defined as $\mathbf{S} = (s_1, s_2, \dots, s_L)$ such that $b_1 = 1, e_L = n, b_{i+1} = e_i{+}1$. This $\mathbf{S}(X) = (\mathrm{X}_1, \mathrm{X}_2, \dots, \mathrm{X}_L)$ now defines the segmentation of sequence $X = (x_1 \dots x_n)$ into segments $((x_1 \dots x_{e_1}), (x_{b_2} \dots x_{e_2}) \dots (x_{b_L} \dots x_{e_L}))$. Likewise, we define $\mathbf{S}(Y)$ for $Y$. From this point forward we represent the segmentation of both sequences, $X$ and $Y$, with $\mathbf{S} = (\mathbf{S}(X), \mathbf{S}(Y)) = ((X_1, X_2, \dots X_{L_X}),$ $(Y_1, Y_2, \dots Y_{L_Y}))$.

Given the segments defined by $\mathbf{S}$, a segmental alignment is a sequence of correspondences $Q = (q_1, q_2 \dots q_T)$ where $q_t = (i_t, j_t), i_t \in \{1, \dots L_X\}, j_t \in \{1, \dots L_y\}$ indicating the matching of segments, such that the following monotonic constraints hold:

$$i_t \in \{i_{t-1}, i_{t-1} + 1\}, j_t \in \{j_{t-1}, j_{t-1} + 1\}. \tag{3}$$

The likelihood of one such fixed alignment $Q$ is defined as

$$P(X, Y|\mathbf{S}, Q, \lambda) = \prod_{t=1}^{T} b_{q_t q_{t-1}}(X, Y) \tag{4}$$

where $\lambda$ encompasses the HMM parameters. Here the likelihood of a match $b_{q_t q_{t-1}}(X,Y)$ is

$$\begin{cases} exp(-\mathcal{D}(\mathrm{X}_{i_t},\mathrm{Y}_{j_t}))\cdot\Psi(|\mathrm{X}_{i_t}|,|\mathrm{Y}_{j_t}|) & i_t=i_{t-1}+1,\\ & j_t=j_{t-1}+1\\ exp(-\sigma_g|\mathrm{X}_{i_t}|) & i_t=i_{t-1}+1,\\ & j_t=j_{t-1}\\ exp(-\sigma_g|\mathrm{Y}_{j_t}|) & i_t=i_{t-1},\\ & j_t=j_{t-1}+1 \end{cases} \tag{5}$$

where $\mathcal{D}(\mathrm{X}_{i_t},\mathrm{Y}_{j_t})$ is the distance between two segments, defined in (2), $\Psi$ specifies the distribution of the corresponding segment lengths, and $\sigma_g$ is a scaling factor. The transition probabilities in the match sequence are defined by the state transition graph in Figure 3 and are denoted by $a$. For instance,

$$a_{q_t q_{t-1} q_{t-2}} = \begin{cases} \delta & , i_{t-1}=i_{t-2}+1, i_t=i_{t-1},\\ & \phantom{,} j_{t-1}=j_{t-2}+1, j_t=j_{t-1}+1\\ \epsilon & , i_{t-1}=i_{t-2}+1, i_t=i_{t-1}+1,\\ & \phantom{,} j_{t-1}=j_{t-2}, j_t=j_{t-1}\\ \tau & , i_{t-1}=i_{t-2}+1, i_t=T,\\ & \phantom{,} j_{t-1}=j_{t-2}+1, j_t=T\\ \text{etc.} \end{cases} \tag{6}$$

with initial transitions, e.g.,

$$a^{(0)}_{q_1} = \begin{cases} \delta & , i_1=0, j_1=1, \quad \text{or} \quad i_1=1, j_1=0\\ 1-2\delta-\tau & , i_1=1, j_1=1\\ \tau & , i_1=0, j_1=0 \end{cases} \tag{7}$$

where $i_1 = 0$ stands for deleting the first segment of $X$ and similarly $j_1 = 0$ denotes deleting the first segment of $Y$. $\Psi$ in (5) can be learned from the data or given as a prior distribution, e.g., uniform. Note that the first case of (5) defines the observation probability of matching two segments (associated with state M in Figure 3) while other cases correspond to gap operations (states I and D).

### 3.1 Inference in SPHMM

An optimal alignment for a fixed segmentation $\mathbf{S}$ can be found as

$$Q^* = \arg\max_Q P(Q|X,Y,\mathbf{S},\lambda) = \\ \arg\max_Q P(X,Y|Q,\mathbf{S},\lambda)P(Q). \tag{8}$$

The prior on $Q$ in (8) can encode traditional band-priors such as the Sakoe-Chiba band. (4)-(8) show that the optimal alignment is the Viterbi path for observing segmented sequences $(X,Y)$.

It is possible to find an optimal segmentation $\mathbf{S}^*$ , together with the optimal alignment, as

$$Q^*,\mathbf{S}^* = \arg\max_{Q,\mathbf{S}} P(\mathbf{S},Q|X,Y,\lambda) = \\ \arg\max_{Q,\mathbf{S}} P(X,Y|\mathbf{S},Q,\lambda)P(\mathbf{S})P(Q), \tag{9}$$

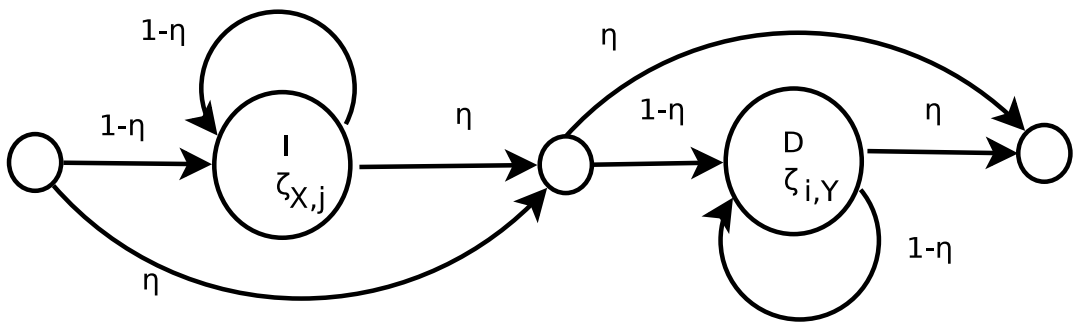

**Fig. 2** Pair-HMM null model.

Although, any informed prior could be used, without loss of generality, we specify uniform prior on **S**. To assert that the alignment likelihood indicates a relationship between the contrasting sequences rather than a random match, one needs to compare the generative sequence likelihood to that of a null model. This null model deletes all segments of one sequence and inserts segments of the contrasting sequence (Figure 2). Therefore, the likelihood of the null model is

$$P(X,Y|\mathbf{S},R) = \left(\eta(1-\eta)^{L_X}\prod_{i=1}^{L_X} exp(-\sigma_g|X_i|)\right) \left(\eta(1-\eta)^{L_Y}\prod_{j=1}^{L_Y} exp(-\sigma_g|Y_i|)\right) \tag{10}$$

where $R$ is the null HMM model with transitions depicted in Figure 2 and observation model similar to (5) (except for the the first equation, which is the likelihood of observing a match between two segments). Thus, assuming that the segmentation priors of the null model and the alternative model are the same, we intend to evaluate

$$Q^*, \mathbf{S}^* = \arg\max_{Q,\mathbf{S}} \frac{P(X,Y|\mathbf{S},Q,\lambda)P(Q)}{P(X,Y|\mathbf{S},R)}. \tag{11}$$

It is possible to evaluate both SPHMM and null model in a single pass over the sequences. In particular, one can assign every match in the SPHMM model to a pair of insertion and a deletion and likewise assign every gap operation to its corresponding insertion or deletion in the null model. Thus, it would be straightforward to formulate reward for match and penalties for opening and extending a gap by expanding (11) with respect to (4) and (10). It helps to observe this formulation in the context of a dynamic programming algorithm for alignment with an affine gap penalty. In particular, for two segments $X_i$ and $Y_j$ the matching reward is

$$r_{mm}(X_i,Y_j) = \frac{1-2\delta-\tau}{(1-\eta)^2} \tag{12}$$

for staying in match state or

$$r_{gm}(X_i,Y_j) = \frac{1-\epsilon-\tau}{(1-\eta)^2} \tag{13}$$

for transitioning from a gap state to match. Consequently, the gap opening penalty for $X_i$ is

$$r_{op}(X_i) = \frac{\delta}{(1-\eta)} \tag{14}$$

and the gap extension penalty is

$$r_{ex}(X_i) = \frac{\epsilon}{(1-\eta)}. \tag{15}$$

By transferring into log-odds ratio the relationship between a Viterbi algorithm and a dynamic programming for alignment is evident. The resulting algorithm is an extension of the best-path algorithm described in Durbin et al (1997)) to segmental model by searching over all permissible segment lengths at each step of the recursion considering the match rewards and gap penalties in (12)-(15). That is, in every state, all possible segments are considered and the segmentation that leads to the highest ratio of posteriors (11) is chosen. To make this procedure computationally tractable one may impose a maximum constraint on the segment length.

**Complexity:**The time complexity of (11) depends both on the lengths of segments in each sequence and the length of the sequences themselves. Given that the number of states is fixed and small, one can prove that the time complexity of the dynamic programming (or marginal matching discussed in Section 3.2) algorithm is $O(l_X l_Y mn)$ where $l_X$ and $l_Y$ are the maximum segment lengths and $n$ and $m$ are the lengths of sequences $X$ and $Y$, respectively. To compute the distance between two segments, one can employ the *summed area table* technique Crow (1984)) to improve the performance. That is, the pairwise distances of all pairs of samples are pre-calculated and the summed area table is constructed. Then within the matching procedure only a few additions are required to compute the distance. Usually, $l_X$ and $l_Y$ are not too long relative to the sequence lengths. Thus, the overall time complexity is typically a small constant factor $l_X l_Y$ away from that of the regular DTW.

### 3.2 Marginal matching likelihood

This subsection introduces an approximation to forward algorithm for segmental pair-HMM. Let us define $\Gamma$ to be the set of all possible segmentations of two sequences $X$ and $Y$ with $m$ and $n$ samples, respectively. Also assume that $\Pi$ is the set of all segmental alignments between $X$ and $Y$. Using the forward algorithm one can estimate the following

$$P(X, Y|\lambda) = \sum_{\mathbf{S}\in\Gamma} \sum_{Q\in\Pi} P(X, Y|Q, \mathbf{S}, \lambda) P(\mathbf{S}) P(Q). \tag{16}$$

We will, again, assume $P(\mathbf{S})$ to be uniform. Computing (16) is not tractable for every possible segmentation. Therefore, we approximate the joint probability of $X$ and $Y$ by explicitly marginalizing over all alignments. That is, we approximate (16) by estimating $P(X, Y|S^*)$ at each step where $S^*$ is a partially optimal segmentation. Specifically, $S^*$ denotes the segments that are optimal only for a partial alignment of the sequences $X$ and $Y$ up to the current step of the algorithm. We use the following recursion to define this approximation.

$$P\left(X_{1:i}, Y_{1:j}|q_t q_{t-1}, \left(S^*(X_{1:(i-k)}), S^*(Y_{1:(j-l)})\right)\lambda\right) = b_{q_t q_{t-1}} \cdot \max_{\mathbf{S}' \in \binom{\Gamma(X_{1:(i-k)}),}{\Gamma(y_{1:(j-l)})}} \sum_{\substack{Q' \in \\ \Pi_{(i-k),(j-l)}}} P\left(X_{1:(i-k)}, Y_{1:(j-l)}|Q', \lambda, \mathbf{S}'\right) \tag{17}$$

where

$$\left(S^*\left(X_{1:i}\right),\ S^*\left(Y_{1:j}\right)\right) = \arg\max_{\mathbf{S}' \in \left(\Gamma(X_{1:i}), \Gamma(Y_{1:j})\right)} \sum_{Q' \in \Pi_{i,j}} P(X_{1:i}, Y_{1:j}|Q', \lambda, \mathbf{S}'). \tag{18}$$

In (17) and (18) $k$ and $l$ are permissible segment lengths for $X$ and $Y$. $\Gamma(.)$ is the set of all segmentations while $S^*(.)$ denotes the approximated segmentation of the given input sequence. $\Pi_{i,j}$ is the set of all possible alignments of $X$ and $Y$ up to $x_i$ and $y_j$. In (17) $q_t q_{t-1}$ defines the current state the same way we defined it in (5). The second term of right hand side of (17) finds the maximum marginalized likelihood over aligning partial sequences given all possible segmentations up to $x_{i-k}, y_{j-l}$. The result of applying this recursive algorithm is the approximated marginalized likelihood of $X$ and $Y$. This is useful in classification problems where one is not necessarily interested in alignment path or optimal segmentation but a reliable likelihood is more desirable. In this paper however, we mainly show the result of the dynamic programming algorithm that arises from (11). The dynamic programming algorithm not only provides us with a likelihood that later can be used as a measure of similarity, but also yields the optimal alignment path and segmentation which is essential to our analysis. We observed superior classification accuracy using the marginal matching algorithm in EEG classification (Section 5).

### 3.3 Learning SPHMM parameters

**Algorithm 1** Learning algorithm for SPHMM.$\#(A \to B)$ denotes the number of transitions from state $A$ to state $B$ decoded by the Viterbi algorithm.

**Initialization**
Randomly initialize $\delta, \epsilon$ and $\tau$. Set $\Psi(i,j)$ to uniform.
**repeat**
  **E-step**: Align training sequences using the Viterbi algorithm described in Section 3
  **M-step**:

1. Re-estimate transition parameters: $\delta = \frac{\#(M\to I)+\#(M\to D)}{2\#(M\to *)}$, $\epsilon = \frac{\#(I\to I)+\#(D\to D)}{\#(I\to *)+\#(D\to *)}$ and $\tau = 1 - 2\delta - \epsilon$.
2. Re-estimate segment length distribution, $\Psi(i,j) = \frac{\#(|\mathrm{X}_{t_X}|=i,|\mathrm{Y}_{t_Y}|=j)}{\#segments}$ $\forall t \in \{1 \dots L_X\}, t_Y \in \{1 \dots L_Y\}$.
3. Tune the parameters using (22) with ($\delta$,$\epsilon$ and $\tau$) as the initial values (project back if needed to respect the feasibility of the starting point)

**until** Convergence.

To learn the parameters of SPHMM one can use a standard expectation maximization algorithm typically used to train HMM parameters Rabiner (1989)). The parameter of the null model cannot be trained using the EM algorithm and must remain constant during training in order to have the consistent reference model. An attractive choice for $\eta$ is the maximum likelihood estimate of (10). That is,

$$\eta = \frac{2}{L_X + L_Y + 2} \tag{19}$$

where $L_X$ and $L_Y$ are number of segments (based on the prior) in each sequence. In our experiments we noticed choosing $\eta$ according to (19) may result into overfitting to the training set in a classification problem and therefore suggest choosing $\eta > 0.5$ in that case.

The standard EM algorithm, does not respect certain constraints that must hold when one designs an alignment algorithm. Those constrains are designed to keep matching reward and gap penalties (Eq. 13-15) within certain bounds. In particular, one would like to have

$$1 < r_{mm}, r_{gm} < z_m, \tag{20}$$

$$z_g < r_{op}, r_{ex} < 1, \tag{21}$$

where $z_m > 1$ and $0 < z_g < 1$ are real numbers. In our experiments we have set $z_m = exp(5)$ and $z_g = exp(-10)$ which provide a reasonable range for learning the parameters.

Maximizing the contribution of matching rewards and gap penalties while satisfying above constraints will lead to solving

$$\begin{aligned}
(\delta^*, \epsilon^*, \tau^*) = \arg\max_{\delta,\epsilon,\tau} (&\hat{c}_{mm} \log(1 - 2\delta - \tau) + \hat{c}_{gm} \log(1 - \epsilon - \tau) \\
&+ \hat{c}_{op} \log(\delta) + \hat{c}_{ex} \log(\epsilon)) && (22) \\
&\text{st.} \\
&2\log(1-\eta) < \log(1 - 2\delta - \tau) < \log(z_m) + 2\log(1-\eta) && (23) \\
&2\log(1-\eta) < \log(1 - \epsilon - \tau) < \log(z_m) + 2\log(1-\eta) && (24) \\
&\log(z_g) + log(1-\eta) < \log(\delta), \log(\epsilon) < \log(1-\eta) && (25) \\
&\log(\tau) < 0 && (26)
\end{aligned}$$

where for $N$ alignments in the training set

$$\hat{c}_{mm} = \frac{\#(M \to M)}{N} \tag{27}$$

$$\hat{c}_{gm} = \frac{\#((IorD) \to M)}{N} \tag{28}$$

$$\hat{c}_{op} = \frac{\#(M \to (IorD))}{N} \tag{29}$$

$$\hat{c}_{ex} = \frac{\#(I \to I) + \#(D \to D)}{N} \tag{30}$$

where $\#(A \to B)$ stands for the number of transitions from state $A$ to $B$. In (22), we have transferred to log-space for numerical stability and used the fact that parameter of the null model ($\eta$) will not be updated. One can transfer (22) into a linear programming by adding $\log(\tau)$ to the objective function and effectively maximize the likelihood of the average Markov model (transitions) under mentioned constraints.

Finally, one can consider the algorithm in Alg.1 for learning the parameters of SPHMM. Note that the inference step is approximated with the dynamic programming resulted from (11). One can incorporate the method described in Section 3.2 to approximate the forward algorithm and use it in a forward-backward learning task (backward algorithm can also be approximated similarly) for estimating the posterior and finally learn the parameters including the distribution of segment lengths. The convergence of the learning algorithm is obvious and provable through the convergence of the EM algorithm. In practice the learning algorithm converges quite fast after a few number of iterations.

## 4 Segmental Matching

In our experiments we observed that during learning SPHMM, the probability of transitioning from match state to gap states can be decreased substantially without significantly affecting the likelihood or alignment path. Given this observation, it is reasonable to expect a single match operation coupled with adaptive segmentation be able to approximate the alignment. Let $\Gamma_m \subset \Gamma$ be the collection of all possible segmentation of $X$ and $Y$ such that: 1) the number of segments is equal in each segmentation, $L = L_X = L_Y$; 2) Corresponding segments are then matched, i.e., the alignment path $Q = (q_1, q_2, \dots q_L)$ where $q_i = (i, i)$. In other words, the alignment is recovered through segmentation. That is,

$$P(X,Y) = \sum_{\mathbf{S} \in \Gamma_m} P(X,Y|\mathbf{S})P(\mathbf{S}) \tag{31}$$

where

$$P(X,Y|\mathbf{S}) = \prod_{t=1}^{L} \exp\left(-\frac{1}{\sigma} D(X_t, Y_t)\right) \Psi(|X_t|, |Y_t|) \tag{32}$$

which is the likelihood of matching two segments in the original SPHMM model. $D(\cdot,\cdot)$ can be any distance metric on sets. Therefore, the joint likelihood of $X$ and $Y$ is maximized by searching over all possible segmentation. That is,

$$P^*(X,Y) = \max_{\mathbf{S} \in \Gamma_m} P(X,Y|\mathbf{S})P(\mathbf{S}) \tag{33}$$

and consequently one may obtain the optimal segmentation as

$$\mathbf{S}^* = \arg \max_{\mathbf{S} \in \Gamma_m} P(X,Y|\mathbf{S})P(\mathbf{S}) \tag{34}$$

where we assume uniform prior on segmentation. A non-uniform prior on segmentation can result into different alignments by favoring longer or shorter segments on different intervals of the sequences. It is possible to compare this model with a random model similar to (10). In that case the prior on segmentation will again cancel out and each matching will be compared to a pair of deletion and insertion.

Removing the two gap operations not only reduces the computational effort incurred by joint segmentation and alignment but also enables one to use bounding methods for particular representations of time-series to further prune the unnecessary computation and speedup the matching. For instance, if the time-series can be locally represented using Bag-of-Words and histogram, often found as a representation in documents or complex video signals, Lampert et al Lampert et al (2009)) have designed bounds on the distance between two segments given a minimum and maximum segment length and their corresponding histograms. We leverage this fact to reduce the computational time of the method proposed in Section 3.

## 4.1 Bounding Histogram Distances

**Bag-of-Words** (BoW): is a popular representation that has been successfully used by researchers Riemenschneider et al (2009)); Chu et al (2012)). In this representation extracted features are clustered into several codewords using a clustering method such as k-means. Similar features described by the same codeword are then counted together and form a histogram for a single or a collection of frames. Therefore, given a histogram map $\phi_{b_i:e_i}(.)$, we denote an $H$-bin histogram of a contiguous segment $b_i : e_i = (b_i, b_i + 1, \ldots, e_i - 1, e_i)$ as $X_{b_i:e_i} = \phi_{b_i:e_i}(V)$ or $X_i$ for short.

Given the maximum segment length $l_{max}$, the minimum segment length $l_{min}$, and two segments of sequence $X$ and $Y$, starting from $b_i$ and $b_j$, respectively, we denote the maximum length segments by $\overline{X}_{b_i} = X_{b_i:b_i+l_{max}}$ and $\overline{Y}_{b_j} = Y_{b_j:b_j+l_{max}}$. Likewise, the minimum length segments are denoted by $\underline{X}_{b_i} = X_{b_i:b_i+l_{min}}$ and $\underline{Y}_{b_j} = Y_{b_j:b_j+l_{min}}$. We are aiming to bound the distance between the histogram features of any possible segment starting from $X_{b_i}$ extending to $X_{b_i+l_{max}}$and $Y_{b_j}$ extending maximally to $Y_{b_i+l_{max}}$. Note that even though we use the same $l_{min}$ and $l_{max}$ for both sequences, it is not a requirement of our method and is used only to simplify the notation. The bin counts of $X_{b_i}$ and $Y_{b_j}$ are bounded as

$$\underline{X}^h_{b_i} \leq X^h_{b_i:b_i+k} \leq \overline{X}^h_{b_i}, (l_{min} \leq k \leq l_{max}) \tag{35}$$

$$\underline{Y}^h_{b_j} \leq Y^h_{b_j:b_j+z} \leq \overline{Y}^h_{b_j}, (l_{min} \leq z \leq l_{max}) \tag{36}$$

where $X^h_.$ and $Y^h_.$ denote the histogram bin $h$.

One can easily extend (35, 36) to normalized histogram noting that $|\underline{X}_{b_i}| \leq X_{b_i:b_i+k} \leq |\overline{X}_{b_i}|$. That is,

$$\frac{\underline{X}^h_{b_i}}{|\overline{X}_{b_i}|} \leq \hat{X}^h_{b_i:b_i+k} \leq \frac{\overline{X}^h_{b_i}}{|\underline{X}_{b_i}|}, (l_{min} \leq k \leq l_{max}) \tag{37}$$

$$\frac{\underline{Y}^h_{b_j}}{|\overline{Y}_{b_j}|} \leq \hat{Y}^h_{b_j:b_j+z} \leq \frac{\overline{Y}^h_{b_j}}{|\underline{Y}_{b_i}|}, (l_{min} \leq z \leq l_{max}) \tag{38}$$

It is straightforward to observe

$$\min(\underline{X}^h_{b_i}, \underline{Y}^h_{b_j}) \leq \min(X^h_{b_i:b_i+k}, Y^h_{b_j:b_j+z}) \leq \min(\overline{X}^h_{b_i}, \overline{Y}^h_{b_j}) \tag{39}$$

$$\max(\underline{X}^h_{b_i}, \underline{Y}^h_{b_j}) \leq \max(X^h_{b_i:b_i+k}, Y^h_{b_j:b_j+z}) \leq \max(\overline{X}^h_{b_i}, \overline{Y}^h_{b_j}) \tag{40}$$

for $l_{min} \leq k, z \leq l_{max}$. Following Chu et al (2012)) one may construct the bounds on popular histogram distances. For completeness of presentation these bounds are included below.

**Bounding $l_1$ distance**: Noting that $|a-b| = \max(a,b) - \min(a,b)$ and a simple reordering of (39, 40) one can observe that

$$\max(\underline{X}^h_{b_i}, \underline{Y}^h_{b_j}) - \min(\overline{X}^h_{b_i}, \overline{Y}^h_{b_j}) \le |X^h_{b_i:b_i+k} - Y^h_{b_j:b_j+z}| \le \max(\overline{X}^h_{b_i}, \overline{Y}^h_{b_j}) - \min(\underline{X}^h_{b_i}, \underline{Y}^h_{b_j}) \quad (41)$$

for $l_{min} \le k, z \le l_{max}$. The bounds on $l_1$ distance are then the summation over all bins. That is,

$$l^{l_1}_b(X_{b_i}, Y_{b_j}, m, l) = \sum_{h=1}^{H} \max(\underline{X}^h_{b_i}, \underline{Y}^h_{b_j}) - \min(\overline{X}^h_{b_i}, \overline{Y}^h_{b_j}) \quad (42)$$

$$u^{l_1}_b(X_{b_i}, Y_{b_j}, m, l) = \sum_{h=1}^{H} \max(\overline{X}^h_{b_i}, \overline{Y}^h_{b_j}) - \min(\underline{X}^h_{b_i}, \underline{Y}^h_{b_j}) \quad (43)$$

and for normalized histograms

$$\hat{l}^{l_1}_b(X_{b_i}, Y_{b_j}, l_{min}, l_{max}) = \sum_{h=1}^{H} \left( \max\left( \frac{\underline{X}^h_{b_i}}{|\overline{X}^h_{b_i}|}, \frac{\underline{Y}^h_{b_j}}{|\overline{Y}^h_{b_j}|} \right) - \min\left( \frac{\overline{X}^h_{b_i}}{|\underline{X}^h_{b_i}|}, \frac{\overline{Y}^h_{b_j}}{|\underline{Y}^h_{b_j}|} \right) \right) \quad (44)$$

$$\hat{u}^{l_1}_b(X_{b_i}, Y_{b_j}, l_{min}, l_{max}) = \sum_{h=1}^{H} \left( \max\left( \frac{\overline{X}^h_{b_i}}{|\underline{X}^h_{b_i}|}, \frac{\overline{Y}^h_{b_j}}{|\underline{Y}^h_{b_j}|} \right) - \min\left( \frac{\overline{X}^h_{b_i}}{|\underline{X}^h_{b_i}|}, \frac{\overline{Y}^h_{b_j}}{|\underline{Y}^h_{b_j}|} \right) \right). \quad (45)$$

Histogram intersection and $\chi^2$ distances can also be derived in the same way.

**Bounding histogram intersection distance**: Histogram intersection distance is defined as

$$d_{\cap}(\phi^H_X, \phi^H_Y) = -\sum_{h=1}^{H} \min(\hat{X}^h, \hat{Y}^h) \quad (46)$$

using (37), (38) the corresponding lower and upper bound is

$$\hat{l}^{\cap}_b(X_{b_i}, Y_{b_j}, l_{min}, l_{max}) = -\sum_{h=1}^{H} \min\left( \frac{\overline{X}^h_{b_i}}{|\underline{X}^h_{b_i}|}, \frac{\overline{Y}^h_{b_j}}{|\underline{Y}^h_{b_j}|} \right) \quad (47)$$

$$\hat{u}^{\cap}_b(X_{b_i}, Y_{b_j}, l_{min}, l_{max}) = -\sum_{h=1}^{H} \min\left( \frac{\underline{X}^h_{b_i}}{|\overline{X}^h_{b_i}|}, \frac{\underline{Y}^h_{b_j}}{|\overline{Y}^h_{b_j}|} \right) \quad (48)$$

**Bounding $\chi^2$ distance**: $\chi^2$ distance is defined as

$$d_{\chi^2}(\phi^H_X, \phi^H_Y) = \sum_{h=1}^{H} \frac{\left(\hat{X}^h - \hat{Y}^h\right)^2}{\hat{X}^h + \hat{Y}^h}. \quad (49)$$

Using the normalized bounds on $l_1$ distance i.e., (44) and (45) one can easily prove

$$\hat{l}_b^{\chi^2}(X_{b_i}, Y_{b_j}, l_{min}, l_{max}) = \sum_{h=1}^{H} \frac{\left(\max(0, \hat{l}_b^{l_1})\right)^2}{\frac{\overline{X}_{b_i}^h}{|\underline{X}_{b_i}^h|} + \frac{\overline{Y}_{b_j}^h}{|\underline{Y}_{b_i}^h|}} \tag{50}$$

$$\hat{u}_b^{\chi^2}(X_{b_i}, Y_{b_j}, l_{min}, l_{max}) = \sum_{h=1}^{H} \frac{(\hat{u}_b^{l_1})^2}{\frac{\underline{X}_{b_i}^h}{|\overline{X}_{b_i}^h|} + \frac{\underline{Y}_{b_j}^h}{|\overline{Y}_{b_j}^h|}} \tag{51}$$

### 4.2 Fast Segmental Matching (Fast-SM)

We propose a recursive algorithm that starts matching from the end of the two sequences. Each segmental match is effectively finding the joint likelihood of $X_i$ and $Y_i$. Within each match we search over all possible segmentations up to the maximum segment length. That is, given $l_{max}$ and $l_{min}$, for $i = L, \dots 1$, $j = L, \dots 1$ and considering uniform prior on segments the likelihood of matching is

$$P(X_{b_i}, Y_{b_j}) = \max_{l_{min} \le k, z \le l_{max}} \exp(-D(X_{b_i-k:b_i}, Y_{b_j-z:b_j}))P(X_{b_i-k-1}, Y_{b_j-z-1}). \tag{52}$$

In other words, (52) is the optimal (maximum) likelihood of matching segments by searching over the likelihood of the last pair of segments in both sequences and all possible segmentation starting from the current point.

We assume that the likelihood of correspondences in the local neighbourhood is approximately constant. Therefore, before executing a recursion to calculate $P(X_{b_i-k-1}, Y_{b_j-z-1})$, we examine the approximated likelihood of the alignment path passing through $(X_{b_i-k}, Y_{b_j-z})$ against the best path found so far. We define $P^*$ as the maximal likelihood calculated for the immediate preceding segment ending in $(X_{b_i-k-1}, Y_{b_j-z-1})$, we have

$$P^* = \max_{\substack{l_{min} \le k' < k \\ l_{min} \le z' < z}} P(X_{b_i-k'-1}, Y_{b_j-z'-1}) \cdot \exp(D(X^*_{b_i-k'-1}, Y^*_{b_j-z'-1})), \tag{53}$$

where $X^*_{b_i-k'-1}$ and $Y^*_{b_j-z'-1}$ denote the best segments extended from $b_i - k' - 1$ and $b_j - z' - 1$, respectively. The second term on the right hand side of (53) cancels out the effect of the last best segment to recover the likelihood of the approximately best alignment up to the neighborhood of $(X_{b_i-k}, Y_{b_j-z})$. Note that all elements required to compute $P^*$ are already calculated and no extra effort is needed to determine it. The bounding is then defined as

$$\tilde{P}(X_{b_i-k-1}, Y_{b_j-z-1}) \le \max_{\substack{l_{min} \le k' < k \\ l_{min} \le z' < z}} P^* \exp(-l_b(X_{b_i-k'-1}, Y_{b_j-z'-1}, l_{min}, l_{max})) \tag{54}$$

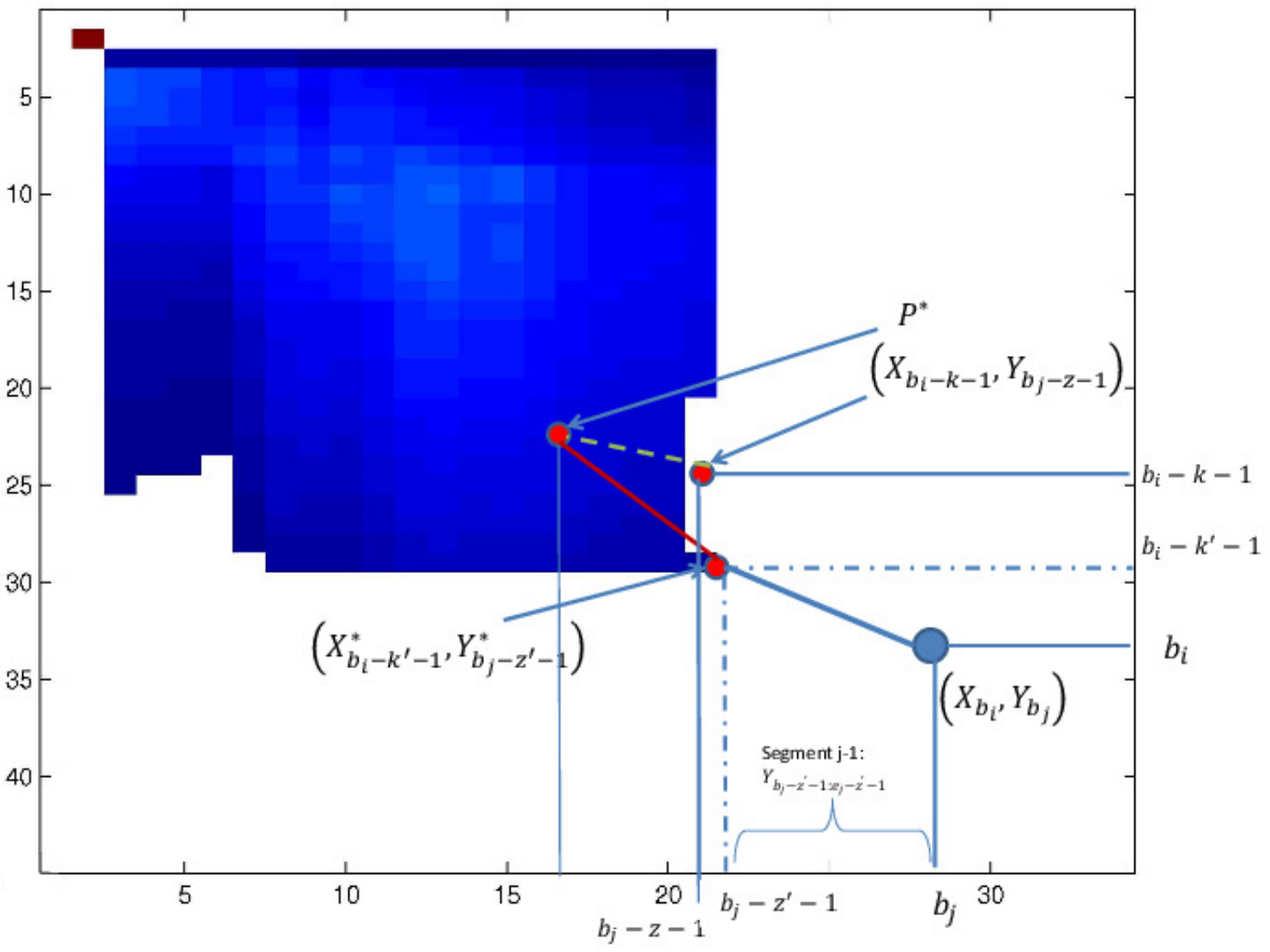

**Fig. 3** Approximate bounding of the likelihood. Axes show the index (time) of contrasting sequences. The shaded area shows the highest alignment likelihood for each correspondence given its optimal segmentation inferred so far. At segment $(X_{b_i}, Y_{b_j})$ we are verifying whether we should expand the new segment to $(X_{b_i-k-1}, Y_{b_j-z-1})$. The best likelihood is achieved by connecting to segment $(X_{b_i-k'-1}, Y_{b_j-z'-1})$ where $l_{min} \leq k' < k$ and $l_{min} \leq z' < z$. Therefore, we can find $P^*$ from which is the likelihood of segmentation up to the end of $(X_{b_i-k'-1}, Y_{b_j-z'-1})$. Then we assume the smoothness on the neighboring likelihood around that point and extend a hypothetical segment from$(X_{b_i-k-1}, Y_{b_j-z-1})$ to it which can be bounded.

where $l_b$ is the corresponding lower bound defined in subsection 4.1. Note that (53) and (54) can be combined for a more efficient implementation. The idea is illustrated in Figure 3. That is, we propose to bound the likelihood of a segment by the the product of the maximal likelihood in its neighbourhood and the upper bound on the likelihood of matching any two segments extended within its boundaries.Therefore, using (54) one can obtain an approximated upper bound on $P(X_{b_i-k-1}, Y_{b_j-z-1})$ and compare it against the best likelihood obtained for the previous segment. We use the term "*approximated upper bound*" since we have made the assumption of smoothness on the local likelihood. If $\tilde{P}(X_{b_i-k-1}, Y_{j-z-1})$ is lower than the best likelihood for the preceding segment obtained so far, we do not expand the recursion and set that corresponding likelihood to its local minimum by

$$P(X_{b_i-k-1}, Y_{b_j-z-1}) = \max_{\substack{l_{min} \leq k' < k \\ l_{min} \leq z' < z}} P^* \exp(-u_b(X_{b_i-k'-1}, Y_{b_j-z'-1}, l_{min}, l_{max})). \quad (55)$$

By setting $P(X_{b_i-k-1}, Y_{b_j-z-1})$ to the minimum likelihood (in the local segmentation sense) we avoid further expansion of this path even if this point is visited again during the segmentation. Note, again, that (53), (54) and (55) can all be combined in the same procedure resulting in a very efficient implementation.

Another technique that contributes to improving the computational performance of our approach stems from the BOW representation. This representation allows one to use the idea of *integral image* Viola and Jones (2004)) to calculate the cumulative sum of the histograms and thus obtain the required segment using a single subtraction operation. That is, if $I$ is a sequence of such cumulative sums one can obtain a segment from $b_i$ to $e_i$ simply by $X_{b_i:e_i} = I_{e_i} - I_{b_i-1}$.

## 5 Experimental Results

In this section we demonstrate the significance of our algorithms through extensive experimentation. We first apply SPHMM on synthetic data to qualitatively assess its performance and also demonstrate its capability in aligning sequences generated by non-causal processes. We then examine our proposed approach on the benchmark data from the UC Riverside "time-series classification page" Keogh et al (2011)). We demonstrate that our algorithm is highly significant compared to DTW, especially in presence of noise. To show that our method is able to deal with non-causal and noisy real-world time-series we also apply it to a publicly available EEG data set, where we demonstrate the benefit of marginal matching algorithm. Finally, we show that SPHMM can significantly improve the performance of activity classification on a subset of HDM05 MoCap data. Segmental matching (SM) and fast segmental matching (Fast-SM) are applied to an activity recognition problem on a publicly available dataset and their superior performance, compared to other algorithms in the literature, is demonstrated.

To measure the performance of the proposed methods and compare with other competing approaches, we use the likelihood reported by each approach as the similarity measure for classification. This is a common way for asserting the goodness of an alignment algorithm quantitatively Ding et al (2008)). Note that the null model is the same for all sequences within a dataset. SM and Fast-SM do not perform as well as SPHMM on the UCR benchmark data as those datasets are specifically chosen for alignment algorithms and gap states are essential to ensure a reliable similarity measurement. They are, however, very competitive to DTW especially when the noise level is high.

Euclidean distance is used as the measure of distance between two samples. We observed that $L_1$ norm can slightly, but not significantly, improve the results in case of excessive noise but we do not include those results. Referring to our discussion in Section 2, employing other distance metrics between sets (such as Hausdorff) resulted in significantly inferior performance especially in noisy data and rendered the alignment of long sequences computationally intractable. Therefore, those results are also omitted from the manuscript. Throughout this section $l_X$ and $l_Y$ denote the maximum allowed lengths of the segments. We have also assumed the scaling parameter of gap operations (5) to be $\sigma_g = 1$. In all experiments the classifier is the baseline 1-Nearest Neighbour (1-NN). We have exclusively used 1-NN to shift the attention from the classifier design to the properties of the similarity measure.

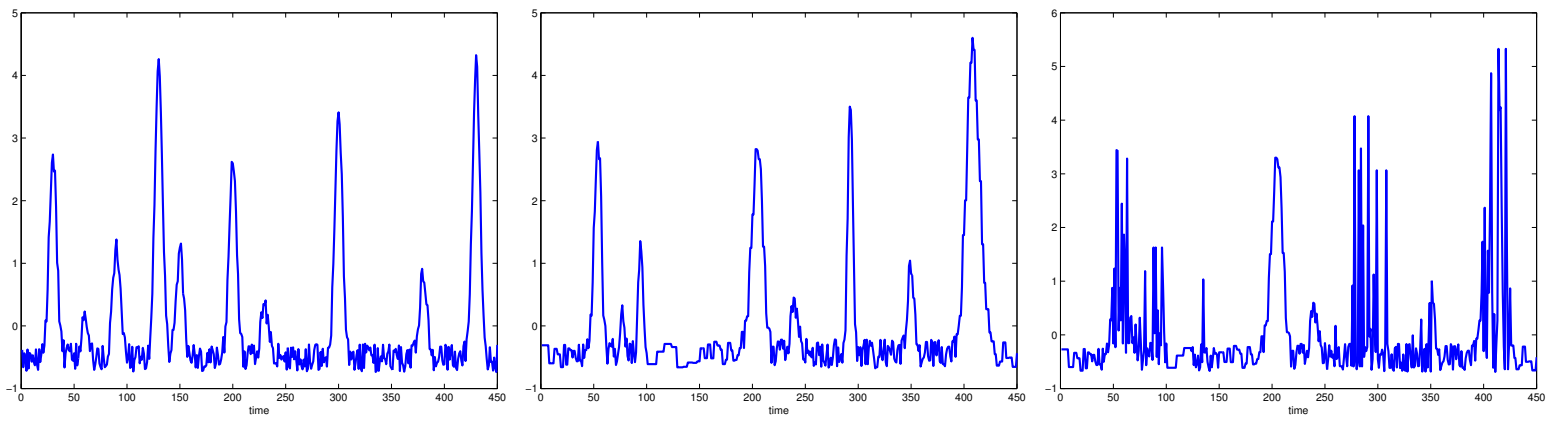

(a) Original (b) Warped (c) Non-causal and warped

**Fig. 4** An instance of a generated sequence and its corresponding warped sequence and its non-causal version.

### 5.1 Synthetic Data I

To demonstrate that our proposed approach can handle non-causal sequences and also have a qualitative comparison with DTW we generated a synthetic dataset and designed the following experiment. 100 sequences are generated from the model

$$T_j(t) = \sum_{i=1}^{10} (\pi_i + \nu_t) \exp\left( (t-\mu)^2 \right) + \omega_t. \tag{56}$$

The time length of all sequences is 450. Peaks in the sequences occur at mean times $\mu = [30, 60, 90, 130, 150, 200, 230, 300, 380, 430]$. The weights are set to $\pi = [7, 1, 3, 10, 3, 6, 1, 8, 3, 10]$ and are corrupted by white independent noise. $\omega_t, \nu_t = N(0, 1)$. We use a monotonic function for the alignment ground truth such that

$$f(t) = \begin{cases} 1 + 0.01 \cdot t^2 & t \leq 100 \\ 310 + 150 \cdot tanh(t/100) & t > 100. \end{cases} \tag{57}$$

To introduce non-causality we add noise to (57) within four intervals such that

$$f^n(t) = \begin{cases} f(t) + N(0, 10) & B_i \leq t \leq E_i \quad \forall i \\ f(t) & otherwise. \end{cases} \tag{58}$$

where $B_i$ and $E_i$ indicate the starting and ending time point of $i^{th}$ non-causal interval. The non-causal time intervals are $[50, 100], [125, 150], [250, 350]$ and $[400, 425]$. For every time-series the contrasting sequence is generated by nearest neighbour interpolation at time points given by (58). A sample of a sequence and its non-causal warped version are shown is Figure 4. SPHMM parameters are learned using Alg. 1 for aligning every sequence and its warped (causal or non-causal) version. We tried segment lengths $l_x = l_y = [50, 100, 150, 200]$.

For a fair comparison with DTW we tried 10 different gap penalties (constant) from 0 to 100, which was applied for every gap operation. Zero gap penalty yielded the best result for DTW. Six of such alignments are depicted in Figure 5. The background is the distance between each sample. The ground truth given by (58) is plotted in red, while the resulting alignment from DTW is drawn in white and that of SPHMM in green. Both axes indicate time and plots are overlaid on the pairwise distance of the two sequences.

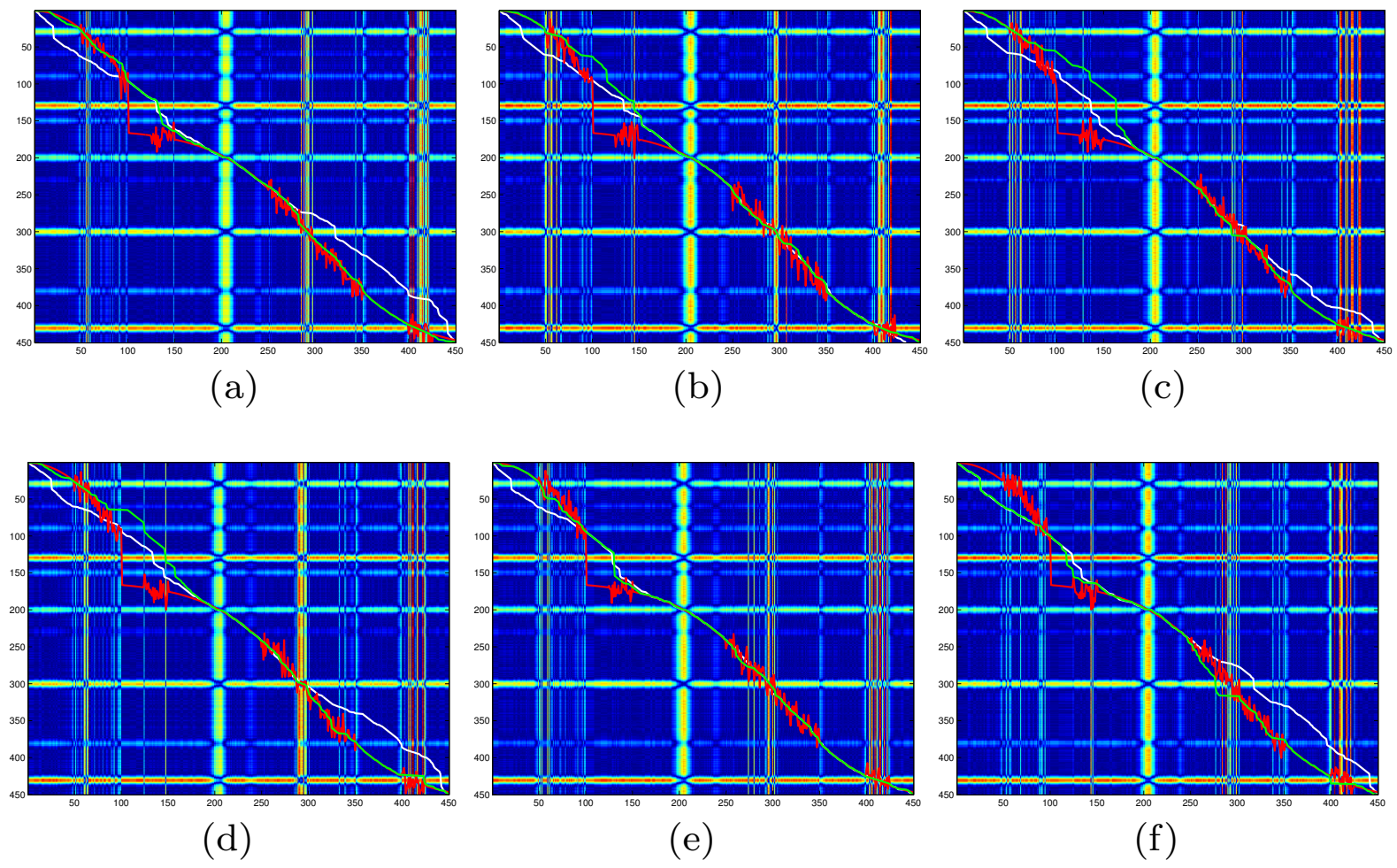

**Fig. 5** Samples of aligning two sequences with non-causal intervals. Each plot depicts the comparison of the ground truth alignment (red) with DTW (white) and SPHMM (Green). The plots show the result for SPHMM with $l_x = l_y = 150$.

It is evident from Figure 5 that SPHMM outperforms DTW in aligning the non-causal time-series. To give a *quantitative* assessment of the goodness of the alignment, the ground truth is compared with reported correspondences by each algorithm. It should be noted that while DTW gives a correspondence for every time-point of the sequence, SPHMM produces segments. These segments are indicated by the starting and ending points. To be able to compare the sequence of segments with ground truth we have used linear interpolation. The goodness measure is the $L_1$ distance of every correspondence from the ground truth. The average $L_1$ distance for DTW over 100 alignments is 8258.8. This value is different for SPHMM for various segment lengths. Namely, the average distance is 7625.5, 5487.1, 5458.5, 5356.0 for $l_x = l_y = [50, 100, 150, 200]$ respectively. It is interesting to note that the distance does not change much for $l_X, l_Y > 100$. The reason is that the largest non-causal interval is 100 time-points long. In many cases the correct segments are extracted except for the second time interval which is located on the valley of the warping function where decoding the correct alignment is difficult for both algorithms.

## 5.2 Synthetic Data II

We also consider the dataset proposed in Shariat and Pavlovic (2011)), where the authors developed an alternative approach to segmental alignment. The dataset consists of sinusoidal and rectangular signals that are embedded into Gaussian noise such that the placement of the signal is also random. Two samples of this dataset are shown in Figure 6. In our original IsoCCA paper we have generated 10 samples from each class and used 1-NN classifier in a leave-one-out setting. We have shown that IsoCCA can achieve 90% accuracy while DTW cannot perform better than 60%. We, however, need to train SPHMM parameters, which is not feasible using a training set derived from 20 sequences. Therefore, we generate 20 more sequences for training the parameters. SPHMM can classify the 20 sequences in test set with 100% accuracy. To assure that the small size of the dataset is not affecting the result we generated 100 sequences and used 5-fold cross-validation setting. We observed that SPHMM is still able to perfectly classify all sequences. This dataset was used in Zakaria et al (2015)); Ye and Keogh (2009)), where the authors show a perfect classification accuracy. Note, however, that their model is not an alignment algorithm and relies on discovering a single motif within each class.

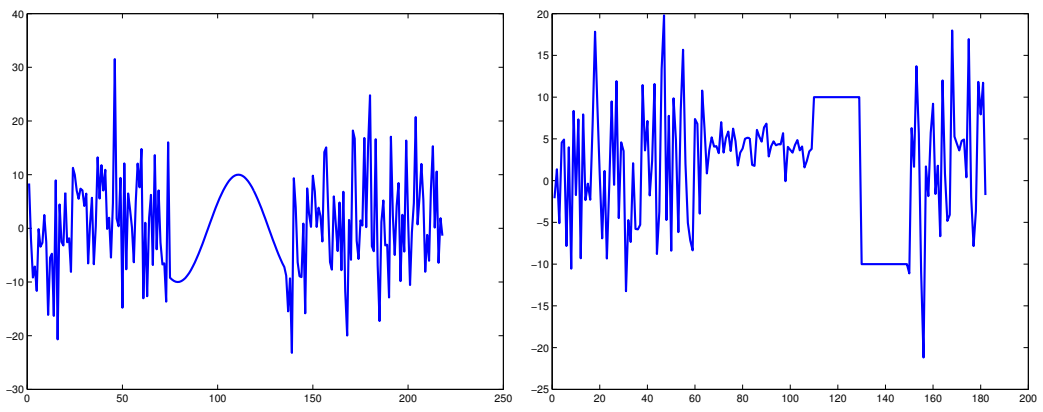

(a) Sinusoidal signal (b) Rectangular signal

**Fig. 6** Synthetic data from Shariat and Pavlovic (2011)).

### 5.3 Benchmark Data

In order to compare our proposed approach to DTW and demonstrate the applicability of our method to general sequences, we tested SPHMM on the entire set of time-series from the UC Riverside time-series repository that contains 45 datasets. The length of time-series in this dataset varies from 60 to 1882. To be able to test the noise resilience of SPHMM, we have added two types of noise to all sequences. The first noise model is the impulse noise. Impulse noise model is very well-known in signal processing community and can model abrupt sensor failure (or other rapid change effects) Abreu et al (1996)). In particular, additive noise process is Gaussian $N(0, \omega\sigma_i)$ where $\sigma_i$ is the standard deviation of the feature $i$ and $\omega$ is the power degree of the noise. We have added the noise to time points chosen uniformly at random, such that the noise does not cover more than 20% of the sequence duration (Figure 7). We conducted the experiment on original data and noisy version of data with $\omega = 1$. For every sequence, we have generated three noisy samples (three noisy sequences) of the corresponding time-series. The algorithms (DTW, PHMM and SPHMM) are then applied to each noisy version of the data and the resultant recognition accuracy results are averaged and reported. The results are shown in table 2.

We compared the proposed approach to DTW and pair-HMM (where no segmentation is applied) with the warping band. To investigate whether DTW with a noise removal pre-processing is superior to SPHMM, we removed the noise using a median filter with two fixed window sizes of 5 and 3 and showed the better recognition rate for each dataset in the DTW-NR column. We have applied the Skao-Chiba band suggested by UCR time-series page to DTW and PHMM. For SPHMM the maximum of the aforementioned band and twice the maximum segment length is chosen as the band to allow SPHMM accommodate up to two segments away from the diagonal of the alignment matrix. The parameters of SPHMM are learned using the method defined in Alg. 1. The segment length distribution however, is not learned and assumed to be uniform. In our experiments we noticed that the model is sensitive to segment length distribution and introducing a non-uniform prior can quickly lead to overfitting. This is due to the fact that the longer segments behave more like outliers. Therefore, it makes sense to use uniform as the segment length distribution. The parameters are not changed for noisy data experiments.

One can see in table 2 that SPHMM is superior or on par with PHMM and DTW in all cases and superior in the original, noise-free setting. However, as soon as the noise is introduced, SPHMM shows a much stronger performance compared to both DTW and PHMM even though PHMM outperforms DTW. One may also notice that even though the median filter noise removal has elevated the recognition rates of DTW (DTW-NR column of impulse noise section in Table 2), it still falls behind SPHMM except for a few cases. The superior performance of DTW-NR in those cases is due to the fact that the window size of median filter accidentally matches the noise spread in one or two noisy versions of those datasets. However, there is no clear way of guessing the correct window size in advance.

To investigate whether the reported results indeed indicate the significance of SPHMM, we have performed Wilcoxon signed rank testDemsar (2006)). In our case for a two-tailed Wilcoxon signed rank test on 45 datasets and $\alpha = .05$, $T = min(R^+, R^-)$ and $z = \frac{T - \frac{1}{4}45\cdot46}{\sqrt{\frac{1}{24}45\cdot46\cdot91}} < -1.95$ was used to assert the significance of the proposed classifier[1]. Table 1 summarizes the results of significance testing. As one can observe SPHMM performs significantly better than other methods in all cases. In the original, noise-free setting, PHMM's performance is not significantly (for $\alpha = 0.05$) superior to that of DTW and both trail the performance of SPHMM. Since the significance of DTW-NR over DTW in the case of noisy data is very much evident, we have not reported this in 1. A standard two tailed Student t-test for asserting the significance of SPHMM results in the same conclusion at 1% significance level for original and 0.1% level for noisy experiments.

**Table 1** Wilcoxon signed rank test for Table 2. ">" stands for "significantly better". Boldface indicates statistically significant relationships.

| | Original | | Impulse Noise | |
|---|---|---|---|---|
| | PHMM $\approx$ DTW | **SPHMM > PHMM** | **DTW-NR > PHMM** | **SPHMM > DTW-NR** |
| $R^+$ | 469 | 590 | 696 | 762 |
| $R^-$ | 396 | 15 | 90 | 228 |
| $z$ | -1.37 | -5.67 | -4.83 | -3.27 |

The average length of the extracted matching segments is approximately 1.08 with a standard deviation of 0.37 in case of noise free data. For the noisy version of the dataset the average length of the matching segments rises to 1.97 with standard deviation of 1.78 indicating that many segments are detected. One has to note that since the chosen data does not result from the random delay processes, detecting many segments of lengths 1, i.e a sample-to-sample matching, is not unexpected.On the other hand, and due to noise (inherent or artificial), it is advantageous to have intermittently extended segments as evident from the reported standard deviation.

[1] $R^+$ ($R^-$) denote the total rank of the datasets where the accuracy of method A is higher (lower) than the accuracy of method B. SeeDemsar (2006)) for details.

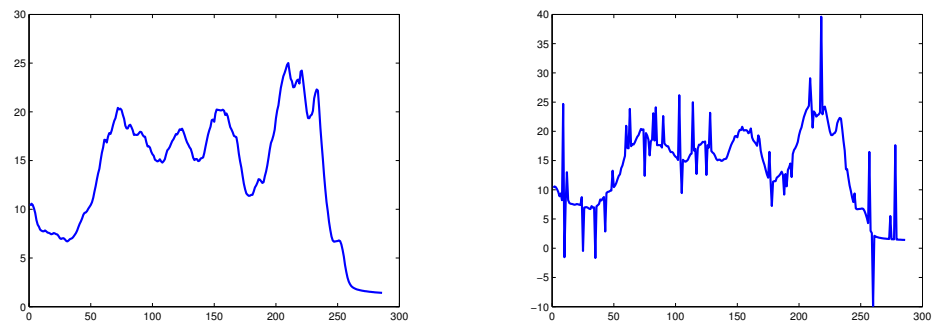

(a) Original (b) $\omega = 100\%$

**Fig. 7** Sample of a sequence from UCR dataset (Coffee) with and without noise.

To demonstrate that our approach is resilient to additive Gaussian noise, we have done the same experiment with the noise spread over the whole span of the signal. Since the noise is more dominant in this case the maximum segment length is increased to 10. We have performed noise-removal using and average filter before applying DTW to make sure that a noise removal with constant window size cannot improve the performance of DTW beyond SPHMM. The average filter window sizes are 10 and 5 and, as we did in the previous experiment, the higher recognition rate is reported. The learned parameters are not changed from the original case. The result is again reported in Table 2. The significance of SPHMM, is obvious and proved by Wilcoxon signed rank test depicted in Table 3. It is interesting to note that noise removal was not able to improve the the performance of DTW and furthermore, in 15 cases has caused a degradation of the performance. This is due to the constant window size and the fact that it does not adapt to the data which is crucial in case of such excessive noise. To assert this conclusion we picked "Trace" and "Adiac" dataset and tried different window sizes for filtering. The result showed significant improvement when the window size is set to 18 for "Trace" and 4 for Adiac. In particular, their accuracy improved to 82.31 and 12.12 for "Trace" and "Adiac", respectively. Another surprising point is that the accuracy results for Beef dataset is higher in noisy case putting the quality of this dataset in doubt (normalization removes this odd behaviour).

We also applied LCSS and EDR algorithm to the noisy data in both impulse and wide-spread Gaussian noise experiments. For brevity, we have not shown those results. Edit-distance based algorithms work significantly better than DTW in case of impulse noise (according to a Wilcoxon signed-rank test) but fall behind the DTW with noise removal pre-processing in that setting. In case of wide-spread additive Gaussian noise they show similar performance to that of DTW. If one applies a noise removal pre-processing before LCSS or EDR, they perform better but again, not better than PHMM.

**Running Time**: Figure 8 depicts the comparison of the average per alignment computation time between DTW and SPHMM when applied to original noiseless data. For short time-series the overhead of computing summed area table is dominant. For longer time-series the computation time is roughly 4 times that of DTW which is much better than the worst case. This is due to the fact that when the algorithm is investigating all segmentations for a correspondence for the first time, it has to find the score of a full alignment for every particular segment. This results in storing the score for every correspondence within all segments originated

**Table 2** UCR time-series classification accuracy in presence of additive Gaussian and impulse noise models.

| | Original | | | Gaussian Noise | | | | Impulse Noise | | | |
|---|---|---|---|---|---|---|---|---|---|---|---|
| | DTW | PHMM | SPHMM | DTW | DTW-NR | PHMM | SPHMM | DTW | DTW-NR | PHMM | SPHMM |
| Lighting7 | 71.23 | 75.34 | **79.45** | 57.99 | 53.42 | 63.01 | **68.03** | 43.51 | 59.10 | 55.01 | **73.94** |
| OSULeaf | 61.98 | 65.7 | **66.12** | 41.32 | 53.99 | 63.77 | **65.01** | 47.11 | 55.55 | 55.15 | **67.18** |
| OliveOil | 83.33 | **86.67** | **86.67** | **37.78** | 27.78 | 35.56 | 35.56 | 28.89 | **51.12** | 28.89 | 32.22 |
| SwedishLeaf | 84.8 | 80.64 | **85.28** | 29.12 | 27.52 | 41.28 | **53.01** | 27.84 | 52.02 | 46.43 | **57.81** |
| Trace | 98 | **100** | **100** | 76.67 | 66.33 | 80.67 | **81.67** | 73.67 | **89.93** | 75.83 | 88.67 |
| Two Patterns | 99.33 | **100** | **100** | 94.66 | 96.24 | 95.36 | **96.85** | 88.22 | 99.85 | 89.86 | **99.96** |
| fish | 82.86 | **86.86** | **86.86** | 35.43 | 33.91 | 36.38 | **38.47** | 34.18 | 60.13 | 60.70 | **71.21** |
| synthetic control | **98.67** | 96.67 | 97.33 | 82.33 | 60.78 | 83.55 | **85.33** | 92.78 | **98.33** | 92.89 | 93.2 |
| wafer | 99.56 | **99.76** | **99.79** | 99.44 | 95.38 | 99.03 | **99.79** | 84.01 | 97.21 | 89.60 | **99.39** |
| yoga | 84.17 | 84.2 | **84.23** | **78.19** | 76.86 | 72.06 | 72.87 | 63.00 | 68.18 | 65.77 | **77.18** |
| 50words | 77.14 | 80 | **80.44** | 29.08 | 70.18 | 70.48 | **71.14** | 57.21 | 74.12 | 74.12 | **77.87** |
| Adiac | 60.61 | **60.87** | **60.87** | 10.66 | 7.33 | 10.91 | **14.41** | 10.20 | 28.17 | 14.59 | **40.04** |
| Beef | **53.33** | **53.33** | **53.33** | 53.33 | 54.44 | **55.55** | **55.55** | 40.00 | 50.00 | 50.00 | **53.33** |
| CBF | 99.67 | **99.89** | **99.89** | 85.78 | 64.71 | 88.11 | **88.74** | 74.35 | 97.33 | 85.93 | **98.01** |
| Coffee | 82.14 | 78.57 | **87** | 65.47 | 70.24 | 60.71 | **87** | 57.14 | 73.81 | 63.22 | **76.78** |
| ECG200 | 88 | **91** | **91** | 85 | 72.67 | 84.33 | **86** | 77.00 | 78.00 | 81.00 | **85.00** |
| FaceAll | **81.72** | 77.51 | 79.59 | 63.89 | 27.97 | 66.31 | **72.25** | 67.89 | 66.84 | 69.05 | **77.20** |
| FaceFour | 89.77 | 89.77 | **92.05** | 84.47 | 73.48 | 87.5 | **90.15** | 52.65 | 80.04 | 68.88 | **89.07** |
| Gun Point | 92 | **98** | **98** | **76.22** | 70.67 | 66.22 | 68.45 | 71.33 | 83.31 | 75.80 | **84.65** |
| Lighting2 | **86.89** | **86.89** | 85.25 | 75.96 | 71.04 | 81.42 | **83.61** | 61.97 | **87.43** | 76.89 | 86.89 |
| ChlorineConcentration | 64.9 | 65 | **66.95** | 42.37 | 38.37 | 45.04 | **52.44** | 38.32 | 50.12 | 41.28 | **53.29** |
| CinC ECG torso | 92.9 | **97.83** | **97.83** | **93.31** | 91.31 | 92.9 | 92.9 | 85.39 | 92.9 | 92.9 | **97.83** |
| Cricket X | **76.15** | 67.69 | **76.15** | 65.3 | 69.3 | 71.41 | **75.11** | 59.06 | **76.15** | 66.24 | 76.05 |
| Cricket Y | 80.51 | 77.95 | **82.82** | 47.52 | 46.52 | 51.51 | **58.21** | 41.98 | 53.82 | 48.51 | **54.19** |
| Cricket Z | **81.79** | 73.33 | 81.54 | 68.37 | 64.37 | 70.67 | **81.55** | 63.83 | **81.79** | 76.76 | 81.54 |
| DiatomSizeReduction | 95.42 | 93.46 | **95.75** | 78.32 | 75.32 | 85.25 | **92.69** | 69.83 | 90.52 | 81.69 | **92.47** |
| ECGFiveDays | 79.67 | 93.73 | **94.08** | 77.93 | 73.93 | **79.67** | **79.67** | 71.48 | 79.67 | 79.67 | **94.08** |
| FacesUCR | 91.27 | 91.85 | **96.88** | 36.94 | 33.94 | 39.27 | **45.36** | 31.41 | 41.27 | 36.29 | **45.04** |
| Haptics | **41.56** | 35.06 | 36.66 | 37.77 | **41.56** | 40.15 | **41.56** | 34.16 | **41.56** | 39.47 | 36.66 |
| InlineSkate | 38.73 | 41.45 | **44.37** | 28.97 | 26.97 | 29.79 | **35.44** | 27.2 | **35.87** | 33.43 | 33.41 |
| ItalyPowerDemand | **95.53** | **95.53** | **95.53** | 78.82 | 80.82 | 85.78 | **89.25** | 72.28 | **95.53** | 82.52 | **95.53** |
| MALLAT | 93.26 | 91.47 | **97.14** | 80.44 | 82.44 | 82.5 | **93.07** | 73.75 | 92.54 | 88.22 | **97.14** |
| MedicalImages | 74.61 | 69.87 | **74.68** | 31.97 | 34.97 | 32.98 | **36.79** | 30.91 | 42.63 | 37.72 | **44.37** |
| MoteStrain | 87.86 | 86.34 | **93.3** | 84.82 | 86.82 | **87.86** | **87.86** | 74.71 | 87.86 | 86.33 | **93.3** |
| NonInvasiveFatalECG Thorax1 | 81.48 | 82.9 | **87.84** | 11.82 | 8.82 | **13.56** | 13.23 | 10.69 | 13.7 | 13.35 | **17.31** |
| NonInvasiveFatalECG Thorax2 | 87.02 | 88.04 | **93.4** | 21.63 | 17.63 | 19.99 | **29.22** | 16.56 | 21.23 | 17.13 | **22.17** |
| SonyAIBORobot Surface | 69.55 | 75.87 | **79.61** | **84.23** | 69.55 | 69.55 | 69.55 | 69.55 | 69.55 | 69.55 | **79.61** |
| SonyAIBORobot SurfaceII | 85.94 | 85.52 | **91.28** | 84.23 | 85.94 | **85.94** | **85.94** | 73.18 | 85.94 | 84.56 | **91.28** |
| StarLightCurves | 86.07 | 84.91 | **90.13** | **89.7** | 86.07 | 86.07 | 86.07 | 82.13 | 86.07 | 86.07 | **90.13** |
| Symbols | 93.77 | 92.06 | **97.88** | 78.39 | 80.39 | 82.32 | **90.77** | 67.9 | 84.04 | 79.46 | **93.89** |
| TwoLeadECG | **85.51** | 74.89 | 79.35 | 62.84 | 62.84 | 65.8 | **73.36** | 59.83 | 74.7 | 67.13 | **79.35** |
| WordsSynonyms | 74.29 | 68.97 | **76.66** | 15 | 12 | 17.94 | **19.79** | 15.57 | 20.96 | 18.99 | **21.84** |
| uWaveGestureLibrary X | 77.44 | 75.1 | **81.18** | 60.6 | 56.6 | 62.42 | **72.85** | 54.81 | 68.26 | 62.33 | **72.36** |
| uWaveGestureLibrary Y | 69.68 | 66.72 | **72.24** | 59.91 | 64.91 | 63.68 | **67.08** | 53.18 | 69.17 | 58.45 | **72.24** |
| uWaveGestureLibrary Z | 67.9 | 65.41 | **70.78** | 48.96 | 47.96 | 55.04 | **56.82** | 43.28 | 56.48 | 49.01 | **58.93** |

**Table 3** Wilcoxon signed rank test for Table 2 additive Gaussian noise section. ">" stands for "significantly better". Boldface indicates statistically significant relationships.

| | DTW > DTW-NR | PHMM > DTW | SPHMM > PHMM |
|---|---|---|---|
| $R^+$ | 649 | 810 | 674 |
| $R^-$ | 341 | 225 | 37 |
| $z$ | -1.99 | -3.30 | -5.42 |

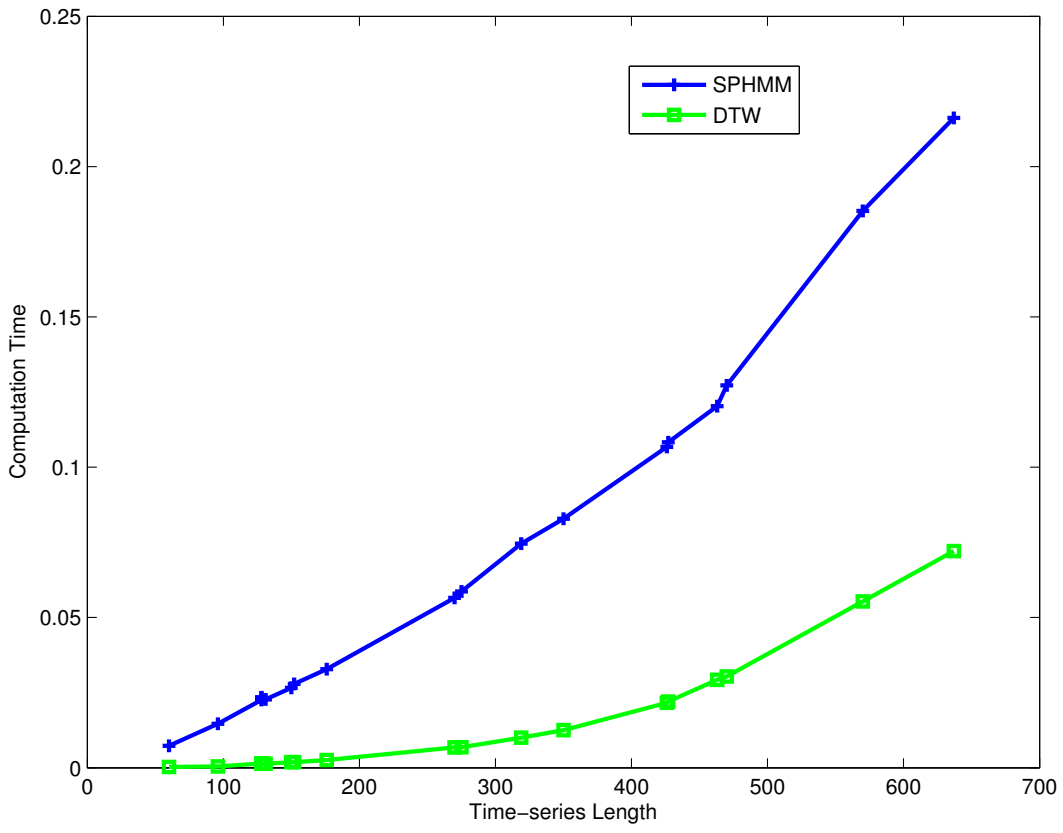

**Fig. 8** Comparison of the average per alignment computation time of SPHMM and DTW on 20 datasets of UCR repository. Vertical axis show the time in seconds

from that correspondence. Therefore, it is not necessary to recompute those values later when investigating the segmentations for neighbouring correspondences (neighbourhood is defined by the maximum segment length). Therefore, it is evident that the algorithm is computationally very efficient given the fact that it significantly outperforms all the rival methods, shows strong robustness against multiple types of noise in addition to producing the joint segmentation and alignment.

### 5.4 EEG Signal Classification

We repeat the experiment on EEG signal classification reported in Shariat and Pavlovic (2012)) to compare the marginal matching algorithm resulted from (17) is the dynamic programming suggested by (11). This experiment also asserts the effectiveness of SPHMM in case of non-causal and noisy real-world time-series. We used the P300 dataset described inHoffmann et al (2005)). Four session are held for each subject. In each session six runs are conducted such that the set of all 6 images is shown at least 20 times to each subject where one of the images is the target in each run. We chose subject 1 and target 2 for our experiment. In each fold of cross-validation we keep one session as training and the remaining three are used as the test set such that every session is used as training once. 1-NN is used as the classifier within a 5-fold cross-validation. We applied the default pre-processing on the data except that we increased the sub-sampling rate to 128 from 32 to acquire longer signals (129 samples). As recommended in the original paper, we only kept 8 channels. The maximum segment length is 20 for both marginal matching and the dynamic programming. Using the dynamic programming an accuracy of 82.64(±1.35) is achieved while using the forward algorithm yielded 84.1(±1.64) which shows a marginal advantage for the marginal matching algorithm.

**Table 4** Confusion matrix of action recognition for SPHMM(in percentage points)

| | DFR | JJack | KRF | KRS | PLF | PRF | Sq | W2S |
|---|---|---|---|---|---|---|---|---|
| DepositFloorR | 65.6 | 0 | 0 | 0 | 6.3 | 3.1 | 0 | 25 |
| JumpingJack | 0 | 98 | 0 | 0 | 0 | 0 | 0 | 2 |
| KickRFront | 0 | 0 | 75.9 | 20.1 | 0 | 0 | 0 | 3.5 |
| KickRSide | 0 | 0 | 21.40 | 71.4 | 0 | 3.6 | 0 | 3.6 |
| PunchLFront | 0 | 0 | 3.6 | 3.6 | 82.1 | 10.7 | 0 | 0 |
| PunchRFront | 0 | 6.7 | 0 | 6.7 | 6.7 | 80 | 0 | 0 |
| Squat | 0 | 0 | 0 | 0 | 0 | 0 | 100 | 0 |
| Walk2Steps | 0 | 3.5 | 3.5 | 0 | 0 | 0 | 0 | 93.1 |

## 5.5 Motion Capture Data

In order to show the effectiveness of our model in a challenging real-world application we performed experiment on HDM05 motion-capture (MoCap) dataset Müller et al (2007)). The actions are usually comprised of several sub-actions. Even plain actions such as walking can be divided into walking with larger or shorter strides at different ends of the line. Therefore, an algorithm that can potentially recover and leverage the subaction segments can outperform alternate approaches. We examine that hypothesis in this experiment.

HDM05 contains MoCap data which consists of 2-3 rotation angles of 29 skeletal joints, resulting in 62 joint angle time series. HDM05 includes 100 classes of action performed by 5 subjects. We choose 8 action classes which are *DepositFloorR, JumpingJack, KickRFront, KickRSide, PunchLFront, PunchRFront, Squat, Walk2Steps*. Sequences are around 300 time-points long and the whole dataset contains 276 sequences in total. We perform 5-fold cross validation and 1-NN is our classifier. Maximum segment length is set to 10. We compare our method against DTW, canonical time warping (CTW) Zhou and de la Torre (2009)) and IsoCCA Shariat and Pavlovic (2011)). SPHMM achieved the highest accuracy, 85.5(±6.18). DTW, CTW and IsoCCA yield 70.1(±5.09), 60.2(±5.1) and 75.1(±6.8) respectively. The significance of SPHMM is evident from the reported results. The confusion matrix for this experiment is shown in Table 4. One can notice that *DepositFloorR* is confused with *Walk2Steps* and *KickRFront* with *KickRSide*. It should be noted that *DepositFloorR* contains the action of walking (one or two steps) right before actual depositing. Also *KickRFront* and *KickRSide* are very much alike. *PunchRFront* is also sometimes confused with *KickRFront*, *KickRSide* and *PunchLFront* where one can perceive that those actions have a lot in common making it difficult to distinguish them correctly in some instances.

## 5.6 UT-Interaction

For typical activities that consist of elementary actions, it may often be the case that the ordering of time points inside the segment ought not to affect the action similarity. For instance, elementary actions performed in opposite directions should still be deemed equally similar as the actions performed in the same direction.

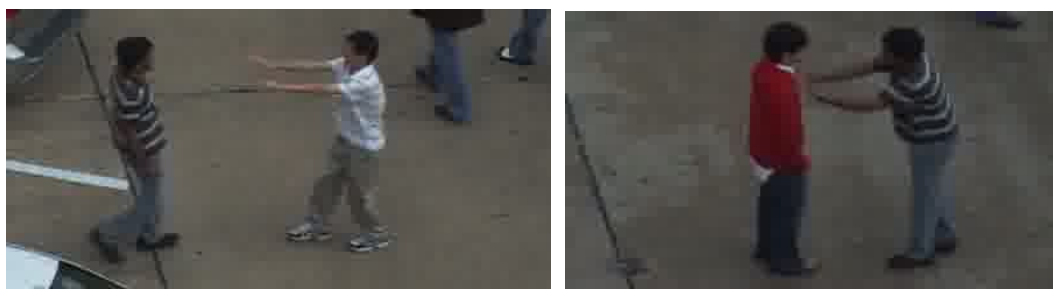

**Fig. 9** Sample frames from UT-interaction dataset #1.

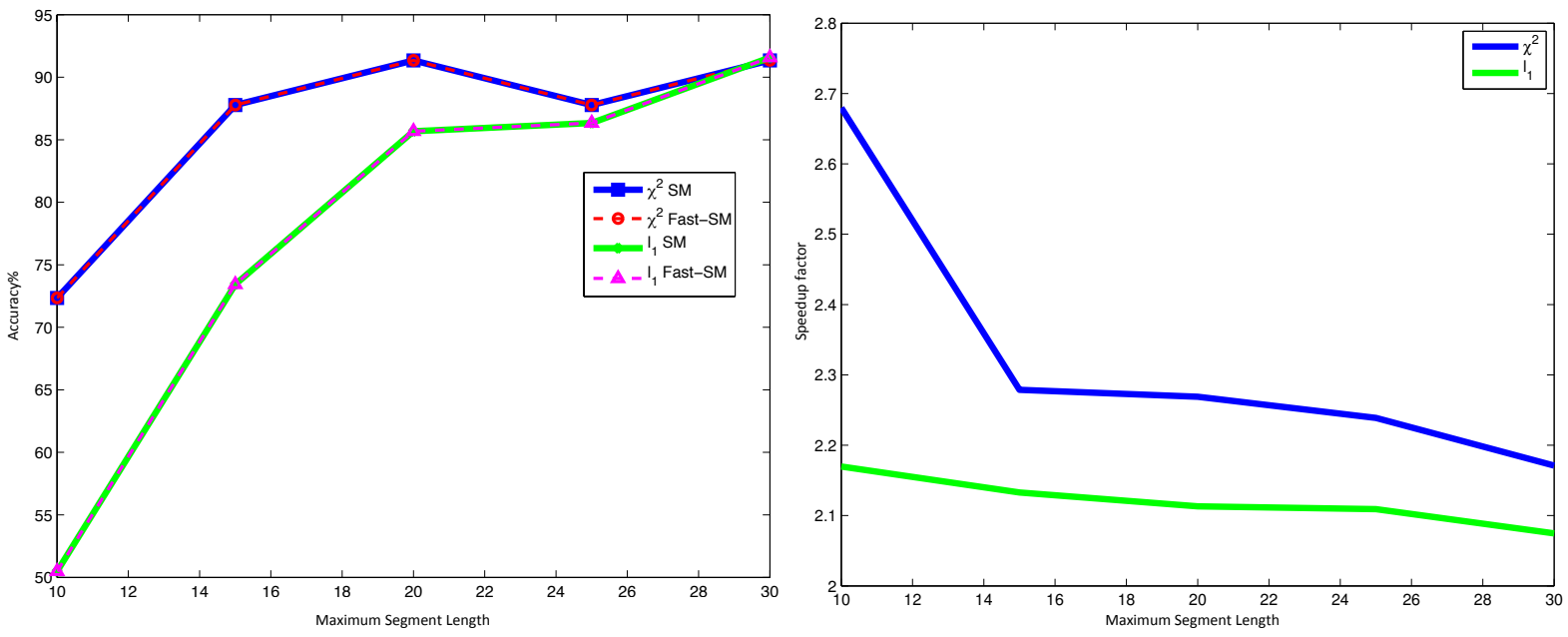

(a) Accuracy (b) Speedup

**Fig. 10** Accuracy and speedup ($\frac{\texttt{SM computation time}}{\texttt{Fast-SM computation time}}$) results for $l_1$ and $\chi^2$ distances as a function of maximum segment length. $l_1$ is depicted as green and $\chi^2$ as blue. Accuracy result of Fast-SM for distance metric is identical to SM.

**Table 5** Recognition rates on UT-interaction dataset #1.

| Method | Accuracy |
|---|---|
| Segmental Match | **91.57%** |
| Dynamic BOW Ryoo (2011)) | 85.0% |
| SVM | 85.0% |
| Voting Waltisberg et al (2010)) | 88.0% |

Therefore, we expect to observe improved performance by applying the segmental matching algorithm to an activity recognition problem.

To apply segmental matching we needed to pick a dataset of reasonable length and complexity so we could try different segmentation lengths and observe how the recognition rate is affected. Therefore, popular action recognition datasets such as KTH Schuldt et al (2004)) or Weizmann Gorelick et al (2007)) datasets were not suitable for our settings because they contain short periodic actions and only a few frames are sufficient for a reliable recognition. Instead, we use the first subset of publicly available UT-interaction dataset containing 10 sequences (60 after segmentation of actions). Within each sequence, six actions, *hand shaking, hugging, kicking, pointing, punching* and *pushing* are performed by 10 different actors. The videos involve camera jitter. Pedestrians are present in the video which makes the recognition more difficult (Figure 9).

We have used spatio-temporal interest points (Cuboids) Dollar et al (2005)) as the descriptors. Then k-means is applied on the resulting features to produce an 800 element codebook.

We use a nearest neighbour classifier to compare with Ryoo (2011)). Leave-one-sequence-out cross-validation by holding one sequence for testing and using the remaining nine for training. Each action in the test set is matched with all training sequences. As a baseline we report the results on SVM using the same feature set and also the results reported in Ryoo (2011)). We have used $l_1$ and $\chi^2$ histogram distances. The results on the $l_1$ distance metric are reported in Table 5. It is evident from the results that our approach significantly outperforms other methods. Using either $l_1$ or $\chi^2$ distance metrics SM and Fast-SM were able to achieve the best result when the maximum segment length was 30. $\chi^2$ achieved the best result even with maximum segment length of 20. We tried different maximum segment lengths, namely, 10,15,20, 25 and 30. Figure 10 illustrates how the resulting accuracy and speedup, gained by bounding the distance (Fast-SM), change as the maximum segment length increases applying $l_1$ and $\chi^2$ histogram distance metrics. It is interesting to note that the recognition rates of Fast-SM and SM are identical in all cases eliciting the fact that the bounding technique and the smoothness assumption on the local likelihoods are in fact effective. In addition, Fast-SM achieves at least a 2-fold speedup compared to SM. As shown in Figure 10(a), $\chi^2$ achieves better results in smaller maximum segment lengths pointing to it as a more suitable measure of distance on segment histograms. Unfortunately, as the maximum segment length increases the bounds on the histogram distances become looser, resulting in reduced speedup. However, one should notice that the shortest sequence is 24 frames long and our final maximum segment length (30) already exceeds this limit. This implies that the model has the option to effectively considers a single BOTW representation as an alternative.

We also applied SPHMM to observe whether a complete alignment model is able to achieve better performance compared to SM and Fast-SM. The result showed that SPHMM cannot advance the recognition rate beyond 91.57% yet, is at least three times slower than SM and four times slower than Fast-SM.

Samples of the discovered segments are depicted in Figure 11. Five activities are illustrated and each segment is separated using a red bar. Only a few frames from each segment is shown. The number of frames shown in each segment is proportional to the length of the segment such that a longer segment is shown with more frames comparing to a shorter segment in the same segmental alignment. An important observation is that the algorithm tends to encapsulate similar relative motions within each segment. For instance, in the 'Hugging' activity (Figure 11(a)), the second and the third segments, which both had the maximum length, encompass the action of hugging. The next segment, shorter in length, contains the pause when the two actors do not move substantially, while the last segment collects the frames corresponding to the actors separating from each other. One can speculate that the second and third segments would merge if the maximum segment length was large enough. However, having larger maximum segment length results in longer running time.

## 6 Conclusion

In this paper we presented a probabilistic model for segmental sequences alignment. We showed that a modified pair-HMM, in conjunction with a proper segment metric, can lead to effective joint segmentation and segmental alignment. Our experimental results showed high accuracy particularly in settings with high levels of noise where DTW loses robustness and, hence, underperforms, even after noise removal pre-processing. Additionally, the invariance to local permutation has enabled our algorithm to perform well on non-causal signals. We also proposed a relaxation of the original model that reduced the computational time. In the particular but common case when histograms are used to represent the time-series we were able to prune the unnecessary computation using bounds on histogram distance metrics.

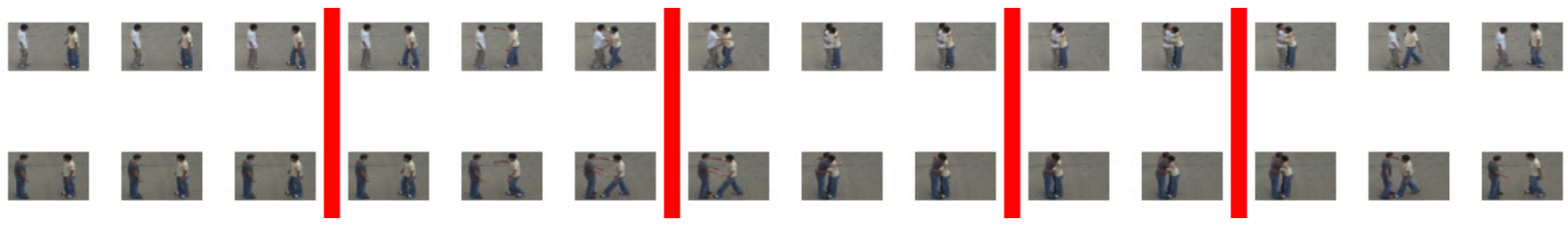

(a) Hugging

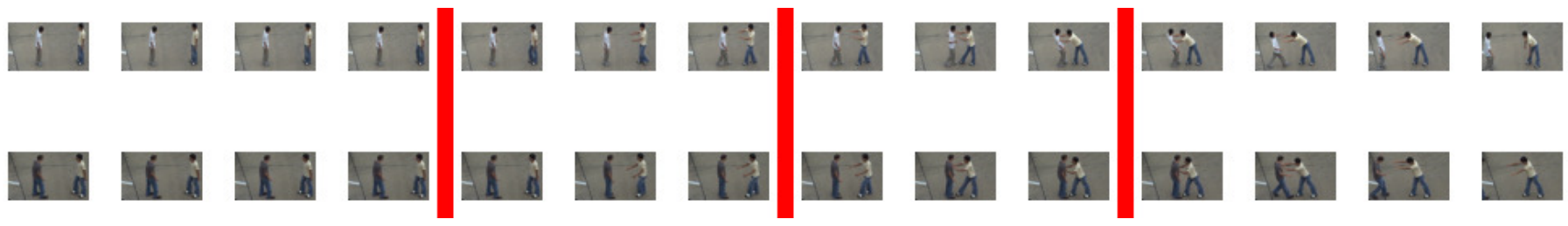

(b) Pushing

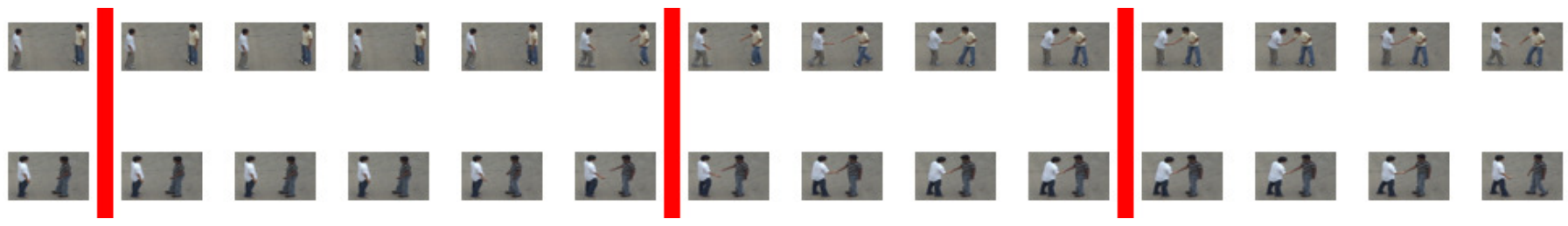

(c) Hand Shaking

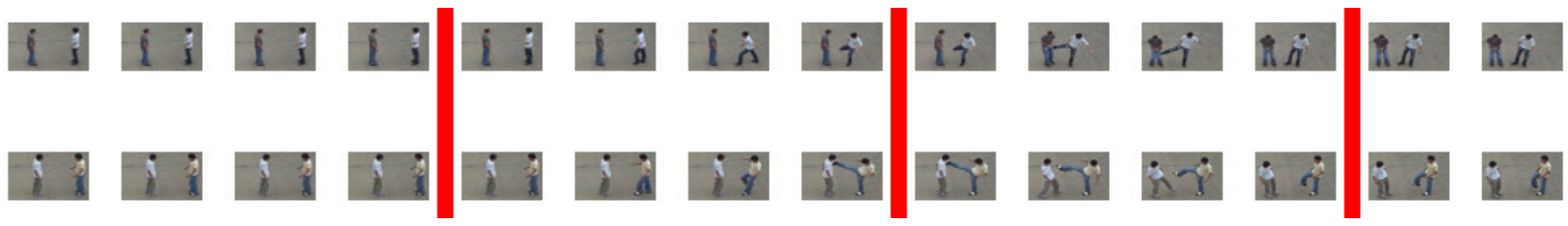

(d) Kicking

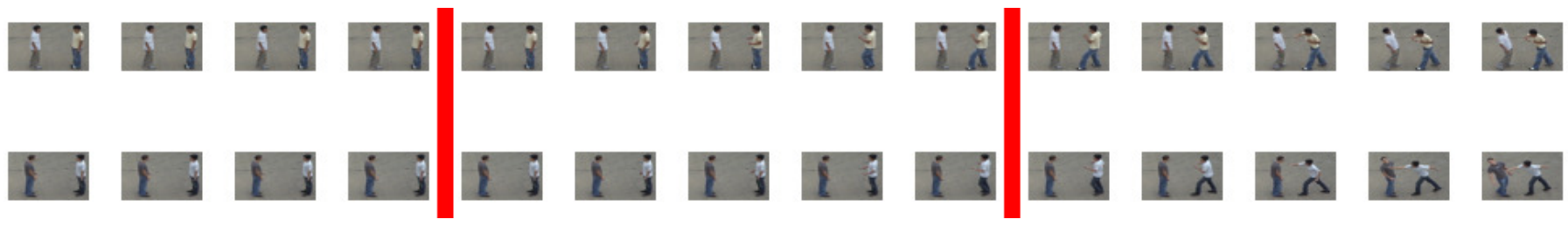

(e) Punching

**Fig. 11** Samples of discovered segments. Segments are separated by red bars. Only a few frames from each segment are shown. The segments and sequences are not necessarily of the same length. The number of frames shown for each segment is increased or decreased for better illustration.

**Author Biographies**

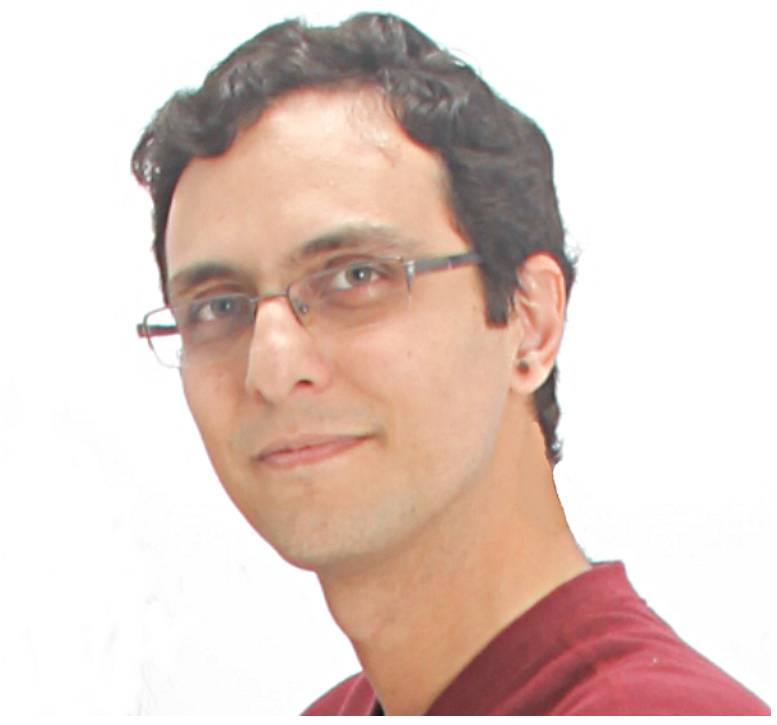

**Shahriar Shariat** received his PhD in computer science from Rutgers University in 2013 and MSc from Sharif University in 2008. Since 2013, he is with applied science team of Turn Inc. as a senior scientist. Shahriar's research interests include time-series analysis and alignment, computer vision and large-scale predictive and statistical learning models. He has many publications in major peer-reviewed venues and has served as a reviewer for several top tier conferences and journals.

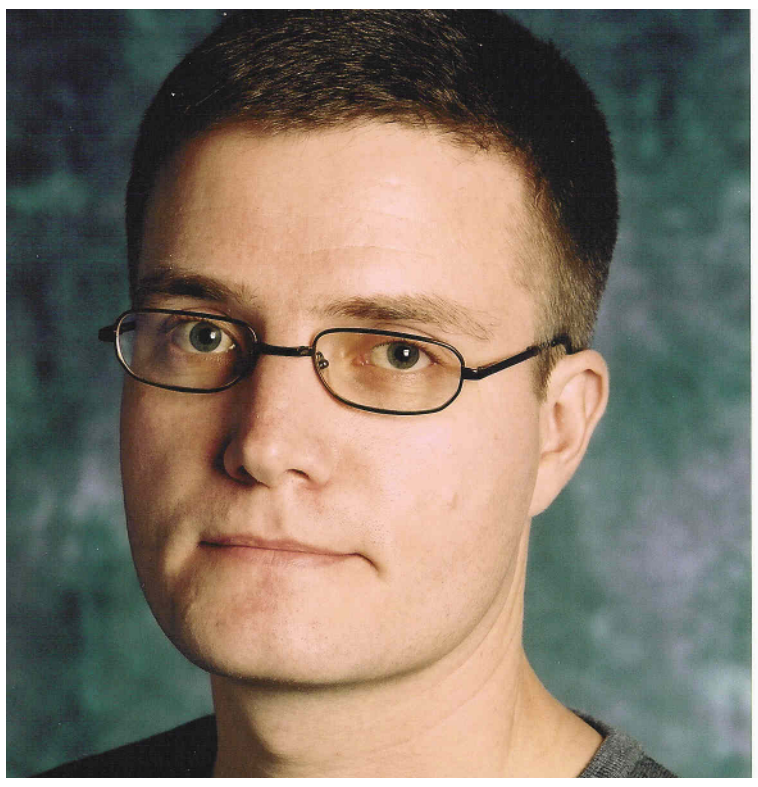

**Vladimir Pavlovic** is an Associate Professor in the Computer Science Department at Rutgers University. He received the PhD in electrical engineering from the University of Illinois in Urbana-Champaign in 1999. From 1999 until 2001 he was a member of research staff at the Cambridge Research Laboratory, Cambridge, MA. Before joining Rutgers in 2002, he held a research professor position in the Bioinformatics Program at Boston University. Vladimir's research interests include probabilistic system modeling, time-series analysis, statistical computer vision and bioinformatics. He has published over 130 peer-reviewed papers in major computer vision, machine learning and pattern recognition journals and conferences.